\documentclass[sigconf]{acmart}
\acmConference[ESEC/FSE 2019]{The 27th ACM Joint European Software Engineering Conference and Symposium on the Foundations of Software Engineering}{26--30 August, 2019}{Tallinn, Estonia}

\usepackage{amsmath,amsfonts}
\usepackage{algorithmic}
\usepackage{graphicx}
\usepackage{textcomp}
\usepackage{xcolor}
\usepackage{tikz}
\usepackage{makecell}
\usepackage{pgfplots}
\usepackage{url}
\usepackage{hyperref}
\usepackage{listings}
\usepackage{xcolor}

\newcommand{\ignore}[1]{}

\definecolor{lbcolor}{rgb}{0.9,0.9,0.9}
\definecolor{darkblue}{rgb}{0, 0.125, 0.576}
\definecolor{gray}{rgb}{0.5,0.5,0.5}
\definecolor{darkgreen}{rgb}{0,0.6,0}

\definecolor{mygreen}{rgb}{0,0.6,0}
\definecolor{mygray}{rgb}{0.5,0.5,0.5}
\definecolor{mymauve}{rgb}{0.58,0,0.82}
\definecolor{dkgreen}{rgb}{0,0.6,0}

\definecolor{lightgray}{rgb}{0.85,0.85,0.85}
\definecolor{lightgreen}{rgb}{0.7,0.9,0.7}
\definecolor{lightblue}{rgb}{0.7,0.7,0.9}
\definecolor{lightred}{rgb}{0.9,0.7,0.7}
\definecolor{OrangeRed}{rgb}{1.0, 0.27, 0.0}

\lstdefinelanguage{diff}{
	frame=single,  
	basicstyle=\ttfamily\scriptsize,
	morecomment=[f][\color{mygreen}]{@@},
	morecomment=[f][\color{darkblue}]{+\ },
	morecomment=[f][\color{OrangeRed}]{-\ },
	keywordstyle=\color{black},
	morekeywords={static, void, if, else, while, struct, int, true, false, unsigned, long, return, goto},   
	framexleftmargin=0.5em,
}

\usetikzlibrary{trees,shapes.misc,shapes.symbols}
\pgfplotsset{compat=1.14}
\usetikzlibrary{positioning}

\def\BibTeX{{\rm B\kern-.05em{\sc i\kern-.025em b}\kern-.08em
		T\kern-.1667em\lower.7ex\hbox{E}\kern-.125emX}}

\setcopyright{none}
\acmDOI{}
\acmISBN{}
\acmConference[]{}{}{}
\acmYear{}
\copyrightyear{}
\acmPrice{}

\begin{document}

\title{Understanding Concurrency Vulnerabilities in Linux Kernel}

\author{Zunchen Huang}
\affiliation{%
	\institution{University of Southern California}
	\city{Los Angeles, CA}
	\country{USA}
}

\author{Shengjian Guo}
\affiliation{%
	\institution{Baidu Security}
	\city{Sunnyvale, CA}
	\country{USA}
}

\author{Meng Wu}
\affiliation{%
	\institution{Virginia Tech}
	\city{Blacksburg, VA}
	\country{USA}
}

\author{Chao Wang}
\affiliation{%
	\institution{University of Southern California}
	\city{Los Angeles, CA}
	\country{USA}
}

\begin{abstract}

While there is a large body of work on analyzing concurrency related
software bugs and developing techniques for detecting and patching
them, little attention has been given to concurrency related security
vulnerabilities.  The two are different in that not all bugs are
vulnerabilities: for a bug to be exploitable, there needs be a way for
attackers to trigger its execution and cause damage, e.g., by
revealing sensitive data or running malicious code.
To fill the gap, we conduct the first empirical study of concurrency
vulnerabilities reported in the Linux operating system in the past ten
years.  We focus on analyzing the confirmed vulnerabilities archived
in the Common Vulnerabilities and Exposures (CVE) database, which are
then categorized into different groups based on bug types, exploit
patterns, and patch strategies adopted by developers.  We use code
snippets to illustrate individual vulnerability types and patch
strategies. We also use statistics to illustrate the entire landscape,
including the percentage of each vulnerability type.
We hope to shed some light on the problem, e.g., concurrency vulnerabilities continue to pose a serious threat
to system security, and it is difficult even for 
kernel developers to analyze and patch them.  Therefore, more efforts
are needed to develop tools and
techniques for analyzing and patching these vulnerabilities.

\end{abstract}
\maketitle

\footnote{October 30, 2019}

\section{Introduction}

\label{sec:intro}

Concurrency bugs are defects caused by unexpected
interleavings of threads or processes that manipulate shared
resources. They are often difficult to reason about
manually because of the nondeterministic nature of thread schedules,
sophisticated data flows, and often astronomically large number of
interleavings.
Over the past decades, there has been sustained interest and steady
progress in developing tools and techniques for detecting and
patching~\cite{Savage1997, Engler2003, Lhee2005, Yu2005, Lu2006,
  Lu2007, Sen2008, LuciaDSC08, Lucia2010, Zhang2010, Zhang2011ConSeq, Sinha2010, EricksonMBO10, TanZP11, Wang11,
  Pratikakis2011, Kasikci2012, Wei2005, Shi2010, CaiC12, CaiWC14, Khoshnood2015, 
  GuoCY17, ZhouSLCL17, CaiCZ17, Wu2015,Kernelthreadsanitizer,
  Choudhary2017, Liu2018, Zhao2018, Yu2018-1, Yu2018-2, Dean2004,
  Jin2012, Lin2018, LiuHuang2018, LiuTZ2014, RungtaMV09, HuangZD13}
concurrency bugs.
Empirical studies~\cite{ChouYCHE01, Lu2008, Fu2018, Stender2008, Paleari2008,
	Wang2017, Warszawski2017, Watson2007, Wei2005, Yang2012, HuangZhang2016, TuLSZ19} also played
an important role.  For example, Lu et al.~\cite{Lu2008} published
their comprehensive study of more than 100 concurrency bugs from
four large open-source applications, which sparked a lot of
research.  Lessons learned from these studies are useful because
they revealed the severity of concurrency bugs in practice, showed
the characteristics of common bug patterns, reviewed the repair
strategies adopted by developers, and more importantly, highlighted
the deficiencies of existing techniques and thus pointed out future
research directions.

However, it remains an open problem \emph{how often concurrency bugs
  lead to security attacks} and, to date, there has been no
comprehensive study on concurrency vulnerabilities.
Concurrency vulnerabilities is a subset of concurrency bugs where
attackers are able to exploit these bugs to cause security damages,
e.g., revealing sensitive data, executing malicious code, disrupting
services, or gaining the root privileges.  Obviously, not all
concurrency bugs are exploitable.  If, for instance, the impact of
triggering a bug remains unobservable to the outside world, it may not
be regarded as a vulnerability.  An example would be the race
condition related to code that creates system logs: here, the bug may
lead to unexpected reordering of log entries, which will annoy users
who inspect the logs, but will not harm the security of the system.
However, there are also concurrency bugs that lead to information
leaks or execution of malicious code; in such cases, the bugs are
regarded as security vulnerabilities as well.

In general, it is not easy to decide \emph{a priori} whether a bug is
exploitable.  A well-known example is a Linux kernel vulnerability,
named \emph{Dirty COW}~\cite{CVE-2016-5195}, which allows attackers to
leverage a race condition bug to change the content
of \emph{read-only} system files.  While the Copy-On-Write (COW)
feature related to this vulnerability was added to the code base long
ago, developers were unaware of it for years and no attack was
revealed to the public either.  Nevertheless, the exploit came to
light ten years later: by leveraging this vulnerability, attackers may
modify system files such as \textit{/etc/passwd} and thus gain root
access to the entire Linux operating system.
Similar incidents have also been reported elsewhere~\cite{Stender2008,
	Yang2012, Warszawski2017, Zhao2018}, indicating that when
concurrency bugs are indeed exploitable, they are often extremely
dangerous.
While security vulnerabilities are routinely discussed in various
online websites, often immediately after they are revealed to the
public, there is a lack of systematic study of them, especially for
concurrency related security vulnerabilities.

To fill the gap, we conduct the first comprehensive study of
concurrency vulnerabilities in Linux kernels~\cite{kernel} together
with the corresponding patches adopted by kernel developers.
Since Linux has open-source code repositories, widely used
issue-tracking systems, excellent documentation, and a community of
avid developers, it has all the ingredients to make a comprehensive
study feasible.
We decide to focus on confirmed vulnerabilities archived in the CVE
database~\cite{NVD}.  This is because deciding whether a concurrency
bug is exploitable is fundamentally challenging, and we do not want
the result of our study biased by our personal opinion.  By focusing
on the CVE entries reported, confirmed, and subsequently patched by
Linux kernel developers, we ensure that the data used in our study are
accurate.

Specifically, for each CVE entry, we manually go through the
descriptions provided by developers as well as the corresponding
parts of the Linux codebase, to understand the nature of the bug.  We
also follow the online discussions, where developers exchange their
opinions regarding the nature of the attack, the proof-of-concept
implementation of the attack, and the various patching strategies.
Only after the CVE entry is fully understood by us, will it be
included in our study.  This ensures that statistics gathered during
our study are accurate.  Since this process of manually sifting
through CVS entries is labor intensive, we cannot afford to go through
each and every entry. Instead, we choose to focus on 101 entries, all
of which have been reported and confirmed in the past ten years, which
happen to represent the latest trend.

We classify these concurrency vulnerabilities into different groups.
First, they are classified based on the bug types.  Then, they are
classified based on the vulnerability types.  We also investigate the
adversarial types, i.e., whether the attack may be launched by a local
user or a remote user.  For each type, we provide a representative
example to illustrate the root cause.  Toward this end, we follow the
presentation style used by Lu et al.~\cite{Lu2008} in their empirical
study of concurrency bugs.  For example, we use code snippets to
illustrate the involved threads, shared variables, and the erroneous
thread interleavings.  We also go beyond what is provided in Lu et
al.~\cite{Lu2008} by explaining how bugs lead to security attacks,
whether the attacks are through information leak, malicious code
execution, service disruption, or allowing the attacker to directly
gain privilege of the system.  Finally, we illustrate the strategies
used by developers to construct patches.

Our study differs from prior work on analyzing concurrency
vulnerabilities in other domains, such as web
applications~\cite{Stender2008} and distributed database
systems~\cite{Paleari2008}. 
It also differs from prior work on analyzing and patching Linux
vulnerabilities~\cite{Wang2017,Wei2005}, including the double-fetch
vulnerability and \emph{Time of Check to Time of Use (TOCTTOU)}
vulnerability.  These previous studies touched upon certain aspects of
concurrency vulnerabilities, but they were not meant to be
comprehensive and therefore did not cover the entire landscape.
Furthermore, vulnerability patterns such as TOCTTOU have become well
understood whereas in this work, our focus is exclusively on CVE
entries reported in the past ten years, which represent the latest
trend.

To capture the entire landscape, e.g., composition of vulnerability
types in real systems and distribution in terms of bug types, exploit
types and repair strategies, we obtain and analyze the corresponding
statistics.  For example, we compute the percentage of individual bug
types, i.e., whether they are \emph{deadlock} or \emph{non-deadlock}
bugs, and in the latter case, whether they are \emph{atomicity
	violations}, \emph{order violations}, or others.  We also provide
statistics on the number of threads and the number of shared variables
involved in the concurrency bugs.  Finally, we provide statistics
related to the repair strategies, as well as the average time taken by
developers to construct the patches.

Based on the study, we reach the following conclusions:
\begin{itemize}
	\item
	Concurrency vulnerabilities pose a serious threat to Linux
	systems as shown by the number of reported vulnerabilities over
	the years, which is increasing steadily, due to the
	ubiquitous multi-core CPUs in today's computer systems.
	\item
	Concurrency vulnerabilities remain difficult to patch.  Although this
	is not a surprise because of the well-known fact that concurrency bugs
	themselves are difficult to fix, the prolonged period between
	reporting and patching dates raises serious concerns.
	\item 
	Since Linux is widely used, e.g., in Android, database systems, and cloud
	computing infrastructures, just to name a few, concurrency
	vulnerabilities also threat the privacy and security of these systems.
	\item 
	When it comes to patching, there is a need to balance the easy of code
	modification and the system performance degradation, which often makes
	it difficult to construct patches.
\end{itemize}
While there have been many tools developed for the purpose of
detecting, replaying and patching concurrency bugs, often times, they
cannot be directly employed by Linux kernel developers to deal with
concurrency vulnerabilities. Indeed, our study of these real
vulnerabilities reported in the past ten years shows that the lives of
developers have not become easier.  Thus, more efforts are needed for
developing automated techniques to aid in the analysis and
repair of concurrency vulnerabilities.

The remainder of this paper is organized as follows.  First, we
motivate our work in Section~\ref{sec:motivation} by pointing out the
main differences between concurrency bugs and concurrency
vulnerabilities.  Then, we explain the methodology used to conduct our
empirical study in Section~\ref{sec:methodology}.  Next, we illustrate
each type of concurrency vulnerabilities in Section~\ref{sec:vulnerability-type}, followed by statistics of these
vulnerabilities in Section~\ref{sec:Vulnerability-result}. We discuss
the repair strategies adopted by developers in
Section~\ref{sec:repair-type}.  We review the related work in
Section~\ref{sec:related} and, finally, give our conclusions in
Section~\ref{sec:conclusion}.

\section{Motivation}
\label{sec:motivation}

In this section, we review the common concurrency bug types and then
use a real example from Linux kernel to illustrate the key ingredients
of a concurrency vulnerability.

\subsection{Concurrency Bugs}
\label{sec:bug}

Following Lu et al.~\cite{Lu2008}, we categorize concurrency bugs into
two groups: \emph{deadlock} and
\emph{non-deadlock} bugs.  Non-deadlock bugs are further
categorized into \emph{atomicity violations},
\emph{order violations} and \emph{others},  depending on what the 
intended execution orders are.  Same as in Lu et al.~\cite{Lu2008}, we
do not focus on \emph{data-races} in our study because data-races
alone may or may not indicate that the corresponding code has a bug;
furthermore, in the CVE database, we find that data-races alone are rarely
associated with any security attacks, although they may appear
together with atomicity or order violations.  In the remainder of this
subsection, we illustrate individual bugs using examples.

\subsubsection{Deadlock}
\label{sec:deadlock}

A deadlock may occur when a thread tries to acquire a shared resource
that has been acquired by another thread.  In some cases, the entity
holding the shared resource may be the thread itself, i.e., a
self-inflicted deadlock.  In Linux, threads often use locks to
coordinate the access of shared resources.  If the coordination is not
implemented correctly, however, a program state may be reached such
that none of the involved threads can make progress.

\begin{figure}
	\vspace{1ex}
	\centering
	\begin{minipage}{.485\linewidth}
\begin{lstlisting}[frame=lrtb]
		/* Thread T1 */    
		spinlock_lock(lk1);
		...             
		spinlock_lock(lk2);  
\end{lstlisting}
	\end{minipage}
	\begin{minipage}{.485\linewidth}
\begin{lstlisting}[firstnumber=6, frame=lrtb]
		/* Thread T2 */
		spinlock_lock(lk2);
		...
		spinlock_lock(lk1);
\end{lstlisting}
	\end{minipage}
	\caption{An example of deadlock.}
	\label{fig:deadlock}
\end{figure}

Figure~\ref{fig:deadlock} shows a scenario where two threads use the
same locks, named \texttt{lk1} and \texttt{1k2}, to enforce critical
sections, where they access shared variables. However, they choose to
acquire the two locks in different orders, which is often a recipe for
disaster.  For example, if thread T1 acquires \texttt{lk1} at the same
time when thread T2 acquires \texttt{lk2}, the program will enter a
state corresponding to (3,8), where 3 and 8 are line numbers of the
two threads.  At this moment, both threads hold the locks needed by
the other; therefore, none of them can make progress. This is what
occurs in some Linux kernel bugs, including the one associated with
CVE-2009-1961~\cite{CVE-2009-1961}, where deadlock may cause a
security attack.

\subsubsection{Atomicity violation}
\label{sec:atom}

Whenever a code region is intended to execute atomically without the
possibility of being interrupted by other threads, developers need to
use dedicated synchronization primitives (e.g., locks or signal/wait)
to enforce the atomic regions.  However, if the intended atomic
regions are not enforced properly, other threads may interleave in
between, thus leading to unexpected behaviors.  Such bugs are called
atomicity violations.

\begin{figure}
	\vspace{1ex}
	\centering
	\begin{minipage}{.485\linewidth}
\begin{lstlisting}[frame=lrtb]
		/* Thread T1 */    
		if(po->rollover!=NULL){ 
		...        
		po->rollover->num = 0;
\end{lstlisting}
	\end{minipage}
	\begin{minipage}{.485\linewidth}
\begin{lstlisting}[frame=lrtb, firstnumber=5]
		/* Thread T2 */
		kfree(po->rollover);
		po->rollover = NULL;
		... 
\end{lstlisting}
	\end{minipage}
	\caption{An example of atomicity violation.}
	\label{fig:atomicity-violation}
\end{figure}

Figure~\ref{fig:atomicity-violation} shows a typical scenario, where
thread T1 first checks if the pointer \texttt{po->rollover} is a
non-null pointer and then uses \texttt{po->rollover} to access the
field named \texttt{num}. However, since thread T2 runs concurrently
with thread T1, it may free the memory pointed to by
\texttt{po->rollover} and then sets the pointer to null. Since 
T1 does not re-check whether the pointer is null immediately before
using it, there will be a null-pointer deference error.  This is what
occurs in \texttt{af\_packet.c} in the Linux kernel
code~\cite{atomicity-violation-af-packet}, where atomicity violation
may crash a process.

\subsubsection{Order violation}
\label{sec:order}

Concurrent operations may need to follow a specific
execution order.  For example, a pointer should never be used prior to
the memory allocation, or after the free.  To make sure such execution
orders are respected, developers must use dedicated synchronization primitives to enforce them.  However, if there is an implementation bug in the thread synchronization code, or developers forget to use synchronizations at all, there will be order violations.

\begin{figure}
	\vspace{1ex}
	\centering
\begin{minipage}{.485\linewidth}
		\begin{lstlisting}[frame=lrtb]
		/* Thread T1 */    
		...                            
		seq_timer_stop(q->timer);
\end{lstlisting}
	\end{minipage}
	\begin{minipage}{.485\linewidth}
\begin{lstlisting}[firstnumber=4, frame=lrtb]
		/* Thread T2 */
		seq_timer_close(q);
		... 
\end{lstlisting}
	\end{minipage}
	\caption{An example of order violation.}
	\label{fig:order-violation}
\end{figure}

Figure~\ref{fig:order-violation} shows a scenario where the
\texttt{stop} operation on \texttt{q->timer} by thread T1 is intended
to occur before the \texttt{close} operation on \texttt{q}. However,
since developers fail to enforce the order, an erroneous interleaving
may reorder the two operations, thus causing a runtime failure.  This
particular order violation, which comes from a file
named \texttt{seq/seq\_queue.c} in the Linux
kernel~\cite{order-violation-seq-queue}, may lead to a security
attack.

\subsection{Concurrency Vulnerabilities}
\label{sec:vul}

Since not all concurrency bugs are vulnerabilities, here, we use an exploitable bug to illustrate the main
ingredients of a vulnerability.  This vulnerability, named \emph{Dirty
	COW}~\cite{CVE-2016-5195}, allows attackers to turn a read-only
mapping of a system file into a writable mapping, thus can be
leveraged to achieve root access of the system.  Here, the
word \emph{COW} stands for \emph{Copy-On-Write}, a mechanism widely
used in operating systems to efficiently implement a duplicate
operation on modifiable resources.  Dirty COW not only affects all
Linux systems but also threats systems built upon Linux, such as
Android.

\subsubsection{Where  is the bug?}
\label{sec:dirtycow}

In Linux, if a memory region is merely duplicated, but not yet
modified, the memory management system will not create a copy;
instead, it allows the original to be used as if it is the new copy.
The actual copy will be created only when modifications must be made
to the copy.  By leveraging this mechanism, the system is able to
increase the speed and reduce the memory consumption significantly.
Furthermore, in Linux, a disk file may be mapped to a memory region,
from which memory read/write operations may be used to access the file
content. Obviously, if a file is read-only, Linux must ensure that
modifications to the memory copy is never written back to the original
file.  Unfortunately, such protection may be skipped by exploiting a
race condition bug associated with Dirty COW.

\begin{figure}
	\vspace{1ex}
	\hspace{-0.08\linewidth}
	\scalebox{1.15}{
		\begin{tikzpicture}
		\tikzstyle{every node}=[font=\small]
		
		\node [text width=1.5cm,align=center] at (-2.0, 1.5) {\tiny{File:\emph{/etc/...}}};
		\draw [thin] (-2.5,1.35) -- (-1.5,1.35);
		\draw[->,thin] (-2.0,1.35) -- (-2.0,0.75) -- (-1.5,0.75);

		\draw [fill=gray,thin,lightgray] (-1.5,0.5) rectangle (0,1);
		\draw [fill=pink,thin,pink] (-1.5,-0.5) rectangle (0,0);
		\draw [thin] (-1.5,-0.6) -- (-1.5,1.8);
		\draw [thin] (0,-0.6) -- (0,1.8);
		\draw [->,thick,blue] (-0.75,0.45) -- (-0.75,0.05);
		\node [text width=1.5cm,align=center] at (-0.75,1.5) {\tiny{Physical Memory}};
		\node [text width=1.5cm,align=center] at (-1.1, 0.25) {\tiny{\emph{Copy}}};
		
		\draw [fill=lightgray,thin,lightgray] (2,1.25) rectangle (3.5,1.75);
		\draw [thin] (2,1) -- (2,1.8);
		\draw [thin] (3.5,1) -- (3.5,1.8);
		\node [text width=1.5cm,align=center] at (4.2, 1.5) {\tiny{Virtual memory} \\ \tiny{in Process P'}};
		\node [rotate=23.3,text width=1.5cm,align=center] at (1, 1.25) {\tiny{\emph{File Read}}};
		\draw[->, thin] (2, 1.5) -- (0,0.75);
		
		\draw [fill=lightgray,thin,lightgray] (2,-0.5) rectangle (3.5,0);
		\draw [thin] (2,-.6) -- (2,0.5);
		\draw [thin] (3.5,-.6) -- (3.5,0.5);
		\node [text width=1.5cm, align=center] at (4.2, -0.25) {\tiny{Virtual memory} \\ \tiny{in Process P}};
		\node [rotate=-24,text width=1.5cm,align=center] at (1, 0.4) {\tiny{\emph{File Read}}};
		\node [text width=1.5cm, align=center] at (1, -0.4) {\tiny{\emph{File write}}};
		\draw[->, thin] (2, -0.2) -- (0,0.7);
		\draw[->, dashed, thin] (2, -0.3) -- (0,-0.3);
		\end{tikzpicture}
	}
	
	\caption{The \emph{Copy-On-Write} Mechanism}
	\label{fig:cow}
\end{figure}
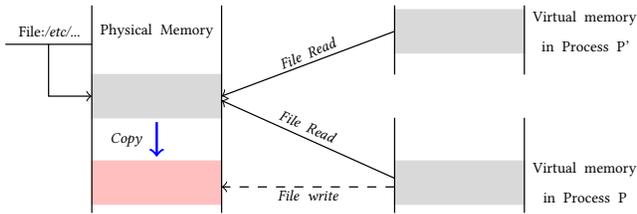

Figure~\ref{fig:cow} shows the COW mechanism in detail. The attacker
may launch a  program that requests the system to map the file
\emph{/etc/xxx} to the system's physical memory; in this figure, $P$ 
and $P$' map their own virtual memory regions to this physical memory
area.  Therefore, both threads can read the file content by reading
their own virtual memory regions.

However, if $P$ tries to write to the file, the kernel will
duplicate the original and hand the new writable memory area to $P$, as
shown by the blue arrow line and the pink area in Figure~\ref{fig:cow}.
Nevertheless, since the virtual memory points to the newly allocated
private memory area (via page table translation), the modification
does not affect the original copy, which is marked as \emph{read
	only}.

Unfortunately, the two-step process of (a) creating a new copy for $P$
and (b) re-mapping the virtual memory of $P$ are not atomic: other
threads or processes may interrupt them, for example, by forcing the  virtual
memory to map back to the original physical memory, just before the
the file write operation.  In such a case, the COW mechanism for
ensuring it never writes a read-only file will be broken.

\subsubsection{How is it exploited?}
\label{sec:attack}

We use a two-threaded program to illustrate how the
vulnerability may be exploited to gain root privilege.  Assume the file \texttt{/etc/passwd} is our target. Normally, a
non-root user cannot alter the file.  However, by running the
two-threaded program as a local user, the attacker can effectively
disable the security check in the COW mechanism, and alter the file.
Since the file contains passwords of all system accounts, by altering
the file, the attacker may create a root account and thus gain control
of the system.

\begin{figure}
	\centering
	\resizebox{0.48\textwidth}{!}{\tt
		\begin{tikzpicture}
		\tikzstyle{every node}=[font=\ttfamily]
		\node[draw, align=left] (usr thread) at (0, 0) {\color{darkgreen}{/* Thread T1 */}\\
			{\tiny 1}  open("/etc/passwd", O\_RDONLY);\\
			{\tiny 2}  map=mmap(...,PROT\_READ,MAP\_PRIVATE);\\
			{\tiny 3}  position=strstr(map,"test:x:1001");\\
			{\tiny 4}  content= "test:x:0000";\\
			{\tiny 5} pthread\_create(...,T2,file\_size);\\
			{\tiny 6}  f=open( "/proc/self/mem",O\_RDWR);\\
			{\tiny 7}  while(1)\{\\
			{\tiny 8}  \textcolor{red!0}{00}lseek(f,position,SEEK\_SET);\\
			{\tiny 9}  \textcolor{red!0}{00}write(f,content,...);\\
			{\tiny 10} \}
		};
		
		\node[draw, align=left] (attacker thread) at (6.5, -0.85) {\color{darkgreen}{/* Thread T2 */}\\
			{\tiny 11}  \textcolor{red!0}{00}while (1) \{\\
			{\tiny 12}  \textcolor{red!0}{00}\textcolor{blue!0}{00}madvise(map,file\_size,\\
			{\tiny 13}  \textcolor{red!0}{00}\textcolor{blue!0}{00}MADV\_DONTNEED);\\
			{\tiny 14}  \textcolor{red!0}{00}\}\\
		};
		\draw[->, dashed, red, thick] (2.2, -1.0) -- (3.6, -0.6);
		\draw[->, dashed, red, thick] (3.7, -0.6) -- (3.7, -1.3);
		\draw[->, dashed, red, thick] (3.7, -1.6) -- (2.0, -1.6);
		\end{tikzpicture}
	}
	\caption{Exploit of the Dirty COW vulnerability}
	\label{fig:dirtycow}
\end{figure}

Figure~\ref{fig:dirtycow} shows the attacker's program in which thread
T1 tries to write a mapped memory copy of the file and thread T2 tries
to discard the copy using an API function named \texttt{madvise()}.
We expect the two threads to interleave and lead to the execution
order outlined by the red dotted lines.

Specifically, thread T1 first opens the file \texttt{etc/passwd} in a
read-only mode at Line 1, and then maps the file to COW memory at Line
2. Next, it looks for the position of the string
\texttt{test:x:1001} where \texttt{test} is the current (non-root) Linux account and the 
digits \texttt{1001} is its UID. Since root account's UID
is \texttt{0000}, the goal of this attack is to simply
replace \texttt{test}'s UID \texttt{1001} with the value \texttt{0000}
in the memory.

Toward this end, T1 first locates the  memory position of
the character sequence (Line 3) and then overwrites it.
However, this operation alone cannot alter \texttt{/etc/passwd}
because the thread is working on a private copy of the file.

This is where T2 helps.  Note that at Line 5, thread T1 creates
thread T2, which then keeps on changing the memory page table using
the MADV\_DONTNEED option. The \texttt{madvise()} system call, which
informs the kernel that the new memory copy is no longer needed and
therefore should be deleted, can take effect between Line 8 and Line
9.  That is, after thread T1 moves the file pointer to the private
copy location, T2 instructs the system to discard the private copy and
map back to the original copy of \texttt{/etc/passwd}.  At this
moment, thread T1 writes (Line 9) the malicious string to the
\texttt{passwd} file, and the account \texttt{test} becomes root.

\subsubsection{How is it patched?}
\label{sec:patchcow}

Due to failures to enforce \emph{atomicity}, the malicious
interleaving of threads T1 and T2 is made possible.  Since both
threads are created by the attacker, and all that is needed to launch
the attack is for the attacker to run a non-root program, the exploit
is both feasible and practical.

The vulnerability may be eliminated in multiple ways.  
For example, one way is to add condition checks immediately
before the second step to confirm that there is indeed privilege to
write to the original memory copy.  
Another way is to introduce extra critical regions, e.g., using
lock/unlock primitives, to make sure the two-step process cannot be
interrupted.
The actual patch adopted by Linux developers is the one that adds the
condition check.  We will explain the repair strategy in detail in
Section \ref{sec:repair-type}.

This vulnerability highlights the danger brought by exploitable
concurrency bugs: when they are indeed exploitable, these bugs often lead
to severe consequences.
Furthermore, they often involve non-trivial thread interactions as
well as subtle functionalities of the operation system.  Thus,
reproducing the attack can be difficult, since it requires not
only orchestration of threads and shared data, but also understanding the
peculiarities of kernel modules.  Due to these reasons, existing
techniques developed for generic concurrency bugs may not be
particularly helpful.
As a community, we need to spend more effort understanding the unique
characteristics of these concurrency vulnerabilities, and then developing
new tools and techniques for analyzing and repairing them.


\begin{table*}
	\centering
	\caption{Our findings of Linux concurrency vulnerability characteristics and their implications.}
	\label{tab:findings}
	\scalebox{.8}{
		\begin{tabular}{|p{.62\linewidth}|p{.58\linewidth}|}
			\hline
			&\\
			\textbf{Findings on Bug Types (Section \ref{sec:Vulnerability-result})}   & \textbf{Implications}  
			\\\hline
			
			(1) All the \emph{non-deadlock} bugs are either atomicity violation or order violation.              
			&  Future vulnerability detection and repair techniques should focus more on these two types.
			\\\hline
			
			(2) Most (80\%) of the \emph{non-deadlock} bugs are atomicity violations.        
			&  Future work on language design, testing, and repair techniques should focus more on enforcing atomic regions.
			\\\hline
			
			(3) Few (5\%) of the vulnerabilities are deadlocks. 
			&  While many kernel bugs may lead to the system hung, few of them are actually caused by deadlocks.
			\\\hline
			
			(4) Almost all (98\%) of the vulnerabilities involve two threads and one variable. 
			&  One may leverage this characteristics to improve the performance of related tools and techniques.
			\\\hline
			\hline
			
			&\\
			\textbf{Findings on Exploit Types (Section \ref{sec:Vulnerability-result})}   & \textbf{Implications}  
			\\\hline
			
			(5) Many (60\%) of the vulnerabilities are caused by Memory Corruption (MC).          
			& Future work on mitigating vulnerabilities should focus more on memory protection in the kernel space.
			\\\hline
			
			(6) Few  (2\%) of the vulnerabilities are caused by Execute Code (EC).         
			& It may be more difficult to execute malicious codes by leveraging concurrency vulnerabilities.
			\\\hline
			
			(7) Many (72\%) of the vulnerabilities can lead to Denial of Service (DoS). 
			& DoS is actually the most common consequence of concurrency vulnerability based attacks.
			\\\hline
			
			(8) About one third (30\%) of the vulnerabilities can lead to Gain Privilege (GP).     
			& GP is obviously dangerous and thus more attention should be devoted to them in the future.
			\\\hline
			
			(9) Few vulnerabilities lead to Gain Information (6\%) and Bypass Something (6\%)       
			& This is based on samples prior to Meltdown/Spectre, which might have changed the situation.
			\\\hline\hline
			
			&\\
			\textbf{Findings on Repair Strategies (Section \ref{sec:repair-type})}   & \textbf{Implications}  
			\\\hline
			
			(10) Almost half (50\%) of the vulnerabilities are patched using \emph{Condition Check}. 
			&  Future work on automated patching tools should devote more attention to Condition Check.
			\\\hline
			
			(11) Few (9\%) of the vulnerabilities are patched using \emph{Code Switch}.  
			&  The correctness of Code Switch is actually difficult to reason about; thus, it may not be easily automated.
			\\\hline
			
		\end{tabular}
	}
\end{table*}

\section{Methodology}
\label{sec:methodology}

In this section, we present the design of our empirical study,
together with a summary of our findings, outlined in
Table \ref{tab:findings}.  Details will be explained in the subsequent
sections. 

\subsection{Vulnerability Sources}


To avoid bias, we do not rely on our personal opinion on whether a bug
is exploitable or not. Instead, we rely on the CVE database, which is
the most trusted source for security professionals.  We focus solely
on Linux kernel vulnerabilities because it is completely open-source
and all related files and documents are easily accessed.  Linux also
has a history of excellent maintenance, supported by a vibrant
community. If there is a vulnerability reported to CVE and confirmed
by kernel developers, it is highly possible that the vulnerability will
be patched eventually. Finally, most of the patches have detailed
explanations and source code on the kernel tree.


We choose 101 samples that are clearly marked as concurrency related
security vulnerabilities by Linux developers.  These samples are
retrieved from the CVE database using the keyword ``race condition'';
that is, they belong to CWE-362-Race Conditions.  We conducted the
search in mid 2018 while restricting the sample to those reported in
the past ten years (2007-2017), which returned 110 samples in total.
We discarded 9 of the 110 samples because they were under
dispute, i.e., it remains unclear whether they are vulnerabilities.

We did not use the keyword ``deadlock'' in search for samples
because the term is loosely used by developers to refer to anything
that may lead to the system hung; in most cases, the cause is
actually unrelated to concurrency.  Since Linux is an operating
system, any bug may cause system hung eventually.  By focusing on
CWE-362-Race Conditions, we guarantee that all vulnerabilities are
indeed concurrency related.


For each of the 101 samples, we collected the following
information: publishing date, which is when the vulnerability is
archived by CVE; and duration between the publishing date and the last
modified date, which is interpreted as the patching duration.
We read all the descriptions, discussions, documentations, and code
snippets, based on which we classified samples by their bug types,
vulnerability types, and repair strategies. 
Here, bug types refer to either deadlock or non-deadlock bugs such as
atomicity and order violations
(Section \ref{sec:bug}). 
We consider the adversarial type, meaning whether the attack can be
launched by a local user, who can run programs on the same computer, or
a remote user.  We also analyzed the bug manifestations, e.g., the
number of threads and variables involved in triggering the bugs.
Finally, we analyzed the repair strategies, which
include \emph{condition check},
\emph{code switch}, \emph{lock strategy} and \emph{design change} (Section \ref{sec:repair-type}).

\subsection{Vulnerability Categories}

Following conventions used by security professionals in
CVE~\cite{CVE-Details}, we classify vulnerabilities into six
categories:
\emph{Memory Corruption (MC)}, \emph{Execute Code (EC)},
\emph{Denial of Service (DoS)}, 
\emph{Gain Privilege (GP)}, \emph{Gain Information (GI)}, and 
\emph{Bypass Something (BS)}.
Note that these categories are not mutually exclusive, e.g., it is
entirely possible that a vulnerability is labeled as both DoS and MC.

Memory Corruption (MC) refers to failures such as use-after-free, null
pointer dereference, double free, and buffer overflow.
Execute Code (EC) refers to execution of malicious code 
injected by the attacker or gathered from the existing code base,
e.g., as in return-to-libc attacks~\cite{Shacham2007}.  
These two types of failures often cause the other four types of
failures.

Denial of Service (DoS) refers to scenarios where legitimate users are
no longer able to access the network, system, local or remote
resources. Sometimes, they may also experience unusually slow system
performance.
Gain Privilege (GP) refers to scenarios where non-root users are able
to gain root access, which is obviously dangerous because it may lead
to all kinds of other security issues. 
Gain Information (GI) refers to scenarios where sensitive information
is revealed to attackers. Although it may seems benign initially, GI
could be the first step of a sequence of destructive attacks. 
Bypass Something (BS) refers to scenarios where security restrictions
such as authentication and file size restriction are skipped
accidentally, which allows attackers to access resources that normally
are not available to them.

\subsection{Threats to Validity}

As with any empirical study, our work has limitations.  First, our
conclusions are valid only for the specific type of vulnerabilities
used in the study. Since we focus on concurrency vulnerabilities in
Linux, the results should not be generalized beyond them.  
Nevertheless, since our sample size is sufficiently large, we expect
that certain characteristics of the concurrency vulnerabilities to be
general enough, and the patterns identified in this paper to be useful
not only for Linux developers, but also other similar systems based
on POSIX threads (PThreads) and written in C.

Since our study used exclusively the confirmed CVE entries, we do not
know how many exploits occurred in practice, but were never reported
to CVE.  Although the CVE database is comprehensive, and often
regarded as the official source by security professionals, it is
entirely possible that some vulnerabilities were patched without ever
being reported to CVE.  Moreover, our data were based on
vulnerabilities reported from 2007 to 2017, which are prior to the
discovery of Meltdown~\cite{Lipp2018meltdown} and
Spectre~\cite{Kocher2018spectre}; therefore, newer types of security
vulnerabilities~\cite{Guo2018,WuW19} may be under-represented or entirely missing.

Linux is written using the C language together with low-level
synchronization primitives such as spin locks.  In other systems and
applications, such as Android and Web applications, the code is often
written using different languages such as Java and JavaScript.
Therefore, the characteristics of concurrency vulnerabilities may be
entirely different.  For example, memory corruption related
vulnerabilities may not be as relevant in Android applications as in
Linux kernel code due to the use of a ``safe'' language (Java) in
Android.

\section{Study of Vulnerabilities}
\label{sec:vulnerability-type}

In this section, we present three case studies of concurrency
vulnerabilities from the National Vulnerability Database
(NVD)\cite{NVD}.  The vulnerabilities correspond to the three bug types:
deadlock, atomicity violation, and order violation. While the deadlock
related vulnerability was revealed almost a decade ago, the other two
were revealed recently.  These three case studies were selected based
on our digest of insightful discussions from the open-source
communities.

\subsection{Deadlock on File Splicing}
\label{sec:dl_pattern}

\begin{figure}	
	\centering
	\resizebox{0.48\textwidth}{!}{
		\begin{tikzpicture}
		\tikzstyle{every node}=[font=\small\ttfamily]
		\node[draw, align=left] (Thread1) at (-2,0.5) {
			{\textcolor{darkgreen}{/* Thread T1 */}}\\
			{\tiny 1} \_file\_splice\_write(pipe\_inode\_info *pipe, \\ 
			\textcolor{red!0}{00}\textcolor{red!0}{00}\textcolor{blue}{struct} file *out)\{\\
			{\tiny 2}\textcolor{red!0}{00} ......\\
			{\tiny 3} \textcolor{red!0}{00}inode\_double\_lock(\textcolor{magenta}{inode},\textcolor{magenta}{pipe->inode});\\
			{\tiny 4} \textcolor{red!0}{00}ret = file\_remove\_suid(out);\\
			{\tiny 5} \textcolor{red!0}{00}{\textcolor{blue}{if}}(likely(!ret))\\
			{\tiny 6} \textcolor{red!0}{0000}ret = \_\_splice\_from\_pipe(\textcolor{magenta}{pipe}, ...);\\
			{\tiny 7} \textcolor{red!0}{00}...... \}
		};
		
		\node[draw, align=left] at (4.3, 0.5) {
			{\textcolor{darkgreen}{/* Thread T2 */}}\\
			{\tiny 10} \_file\_splice\_write( \\
			\textcolor{red!0}{00}\textcolor{red!0}{00}pipe\_inode\_info *pipe, \\
			\textcolor{red!0}{00}\textcolor{red!0}{00}\textcolor{blue}{struct} file *out) \{\\
			{\tiny 11} \textcolor{red!0}{00}......\\
			{\tiny 12} \textcolor{red!0}{00}inode\_double\_lock(\textcolor{magenta}{inode},\\ \textcolor{red!0}{000000}\textcolor{magenta}{pipe->inode});\\
			{\tiny 13} \textcolor{red!0}{00}......\\
			{\tiny 14} \}
		};
		\draw [->, dashed, red, thick] (1.4, 0.6) -- (1.4, -0.9);
		\draw [->, dashed, red, thick] (1.4, -0.85) -- (2.35, 0);

		\node[draw, align=left,text width=12.15cm] at (0.455,-3.5) {\\
			{\tiny 15} inode\_double\_lock(\textcolor{blue}{struct} inode *\textcolor{magenta}{inode1},  \textcolor{blue}{struct} inode *\textcolor{magenta}{inode2})\{\\
			{\tiny 16} \textcolor{red!0}{00}...... \\
			{\tiny 17} \textcolor{red!0}{00}{\textcolor{blue}{if}}(\textcolor{magenta}{inode1} < \textcolor{magenta}{inode2}) \{\\
			{\tiny 18} \textcolor{red!0}{0000}mutex\_lock\_nested(\&\textcolor{magenta}{inode1}->i\_mutex, I\_MUTEX\_PARENT);\\
			{\tiny 19} \textcolor{red!0}{0000}mutex\_lock\_nested(\&\textcolor{magenta}{inode2}->i\_mutex, I\_MUTEX\_CHILD);\\
			{\tiny 20}\textcolor{red!0}{00}\} {\textcolor{blue}{else}} \{\\
			{\tiny 21}\textcolor{red!0}{0000}mutex\_lock\_nested(\&\textcolor{magenta}{inode2}->i\_mutex, I\_MUTEX\_PARENT);\\
			{\tiny 22}\textcolor{red!0}{0000}mutex\_lock\_nested(\&\textcolor{magenta}{inode1}->i\_mutex, I\_MUTEX\_CHILD);\\
			{\tiny 23}\textcolor{red!0}{00}\} \\
			{\tiny 24}\}
		};
		\end{tikzpicture}
	}
	\caption{Deadlock-related functions in fs/splice.c}
	\label{fig:splice}
\end{figure}

Our first case study is on a vulnerability caused by deadlock.  The
deadlock is related to the Linux file splicing.  It occurs when 
two or more threads try to access the same files using lock-based
synchronization.

Recall that the kernel code may write to a file through piped steps,
where each pipe has a \emph{read} end and a \emph{write} end. Data
coming from the \emph{read} end is redirected to the next step by
the \emph{write} end and, finally, is written to the file. As shown in
CVE-2009-1961~\cite{CVE-2009-1961}, due to an implementation bug,
concurrent calls to file splicing with pipe contention may lead to security
attacks.

Figure \ref{fig:splice} shows the related code snippet. The upper-left
function, named \texttt{\_file\_splice\_write}, splices data from a
pipe to a file, whereas the upper-right is a more concise version of
the function.  The third function, named
\texttt{inode\_double\_lock}, is meant to lock two input inodes in 
the \emph{parent-child} order.  Specifically, the file-splicing
operation tries to lock both the pipe and the destination file (Line
3), and then moves data from the pipe to the file inside the
function \texttt{\_\_splice\_from\_pipe} (Line 6).  However, due to a
design error, there may be a deadlock when multiple threads use
different locking orders on the same pipe and file.

Assume \texttt{T1} and \texttt{T2} execute the same
\texttt{\_file\_splice\_write()} over  the same pipe and file, as shown in 
Figure~\ref{fig:splice}.  Following the dotted line, T1 locks both
pipe and file using \texttt{inode\_double\_lock()} (Line 3) before
transferring data in \texttt{\_\_splice\_from\_pipe()} (Line 6).
However, if the pipe is not ready, T1 will have to release the lock
on \texttt{pipe->inode} and go to sleep.
Next, T2 tries to lock the same \texttt{pipe->inode}
and \texttt{inode} in \texttt{inode\_double\_lock()} (Line 12).  At
this moment, T2 can lock \texttt{pipe->inode} since it has been
released by T1, but can never lock \texttt{inode} because it is still
held by T1.

More seriously, the deadlock may be triggered by a (malicious) program
with low privilege, and yet the deadlock itself will force the system
to deny services to file creation and removal to all users, thus
resulting in a DoS attack.

Furthermore, this particular deadlock pattern is actually quite common
in Linux kernel.  Three out of the five deadlock issues in our CVE
entries have similar problems.

\subsection{Atomicity Violation in File Descriptor}
\label{sec:av_pattern}

\begin{figure}	
	\centering
	\resizebox{0.48\textwidth}{!}{
		\begin{tikzpicture}
		\tikzstyle{every node}=[font=\footnotesize\ttfamily]
		\node[draw, align=left,text width=4.8cm] (timerfd) at (-0.2,1.55) {\\
			{\tiny 1} {\color{blue}{struct}} timerfd\_ctx \{\\
			{\tiny 2} \textcolor{red!0}{00} ......\\
			{\tiny 3} \textcolor{red!0}{00} {\color{blue}{struct}} list\_head clist;\\
			{\tiny 4} \textcolor{red!0}{00} {\color{blue}{bool}} might\_cancel; \}\\
		};
		\node[draw, align=left,text width=4.8cm] (usr thread) at (-0.2,-0.8) {\\
			{\textcolor{darkgreen}{/* Thread T1 */}}\\
			{\tiny 5} \_remove (timerdf\_ctx \textcolor{magenta}{*ctx}) \{\\
			{\tiny 6} \textcolor{red!0}{00}{\textcolor{blue}{if}}(\textcolor{magenta}{ctx->might\_cancel})\{\\
			{\tiny 7} \textcolor{red!0}{000}\textcolor{magenta}{ctx->might\_cancel} = false;\\
			{\tiny 8} \textcolor{red!0}{000}spin\_lock(\&cancel\_lock);\\
			{\tiny 9} \textcolor{red!0}{000}list\_del\_rcu(\textcolor{magenta}{\&ctx->clist});\\
			{\tiny 10} \textcolor{red!0}{00}spin\_unlock(\&cancel\_lock);\\
			{\tiny 11} \textcolor{red!0}{0}\}\}\\
		};
		\node[draw, align=left,text width=4.8cm] (attacker thread) at (5.2, -0.0) {\\
			{\textcolor{darkgreen}{/* Thread T2 */}}\\
			{\tiny 12} \_setup (timerfd\_ctx \textcolor{magenta}{*ctx}, ...)\{\\
			{\tiny 13} \textcolor{red!0}{00}......\\ 
			{\tiny 14} \textcolor{red!0}{00}{\textcolor{blue}{if}}(!\textcolor{magenta}{ctx->might\_cancel})\{\\
			{\tiny 15} \textcolor{red!0}{0000}\textcolor{magenta}{ctx->might\_cancel} = true;\\
			{\tiny 16} \textcolor{red!0}{0}\textcolor{red!0}{000}spin\_lock(\&cancel\_lock);\\
			{\tiny 17} \textcolor{red!0}{0000}list\_add\_rcu(\textcolor{magenta}{\&ctx->clist},\\
			\textcolor{red!0}{000000}\&cancel\_list);\\
			{\tiny 18} \textcolor{red!0}{0000}spin\_unlock(\&cancel\_lock);\\
			{\tiny 19} \textcolor{red!0}{00}\}{\textcolor{blue}{else if}}(\textcolor{magenta}{ctx->might\_cancel})\{\\
			{\tiny 20} \textcolor{red!0}{0000}\_remove({\textcolor{magenta}{ctx}});\\
			{\tiny 21} \textcolor{red!0}{00}\}\\
			{\tiny 22} \textcolor{red!0}{00}......\}};
		\draw [->, dashed, red, thick] (2.8,-1) -- (2.1, -0.1);
		\draw [->, dashed, red, thick] (2.1, -0.3) -- (2.1, -1.6);
		\draw [->, dashed, red, thick] (2.2, -1.55) -- (2.7, -1);
		\end{tikzpicture}
	}
	\caption{The \emph{Timer} file descriptor and related operations}
	\label{fig:timerfd}
\end{figure}

In this section, we study how atomicity violation may cause a security
vulnerability in the kernel \emph{timer} object, and how it may be
exploited to launch an attack.

In a Linux kernel, developers can access the
system's \emph{timer} through a file descriptor. The typical use of
\emph{timer} is to monitor the expiration of a running
process. CVE-2017-10661~\cite{CVE-2017-10661} describes a
vulnerability that can be triggered by an attacker running multiple
threads on the same \emph{timer} file descriptor.

Figure~\ref{fig:timerfd} shows the code snippet where both kernel
APIs, \texttt{\_remove} and \texttt{\_setup}, take the file descriptor
pointer (\texttt{*ctx}) as a parameter.  The descriptor, of the
type \texttt{timerdf\_ctx}, is a C structure containing a Boolean
variable named \texttt{might\_cancel} (Line 4), indicating if the
linked list entry \texttt{clist} (Line 3) may be removed from its
host.

The function \texttt{\_remove()} is meant to delete the
node \texttt{clist}.  If \texttt{ctx->might\_cancel} is \emph{true},
then \texttt{\_remove()} changes it to \emph{false} (Line 7) and
removes \texttt{clist} (Line 9).  The function
\texttt{\_setup()} calls \texttt{\_remove()} (Line 20) if its 
\texttt{ctx->might\_cancel} is \emph{true}; otherwise, it sets \texttt{ctx}'s 
\texttt{might\_cancel} to \emph{true} (Line 15) and then inserts 
\texttt{clist} to the linked list (Line 17).

In the ideal case, \texttt{\_setup()} and \texttt{\_remove()} work
cooperatively: the former updates the flag and inserts the node while
the latter resets the flag and removes the inserted node.  However,
the implementation of these two APIs fails to enforce the sequence of
operations on \texttt{ctx} as atomic.

Therefore, an attacker can leverage this bug to trigger a memory
corruption, e.g., by running the following two-threaded program.

\begin{figure}[H]
	\vspace{-2ex}
	\centering
	\begin{minipage}{0.95\linewidth}
		\begin{lstlisting}[numbers=none, frame=single]
		/* Thread T1*/                      /* Thread T2*/
		while(1){_remove(&ctx);}            while(1){_setup(&ctx);}
		\end{lstlisting}
	\end{minipage}
	\vspace{-3ex}
\end{figure}

\noindent
The root cause of the vulnerability is that no locks are used to
protect the \emph{timer} descriptor \texttt{ctx},  hence its field
\texttt{might\_cancel} can be improperly altered, e.g., following
the execution order shown by the red dotted line in
Figure~\ref{fig:timerfd}.  That is, \texttt{T2} runs first and
switches to \texttt{T1} immediately before Line 20 in its second
invocation of \texttt{\_setup()}. At this moment, the
flag \texttt{might\_cancel} is \emph{true} because of the update
in \texttt{T2}'s first invocation of
\texttt{\_setup()} (Line 15 in Figure~\ref{fig:timerfd}).  Then, 
assume that \texttt{T1} runs \texttt{\_remove()} once to delete the
list node. After that, the execution resumes to \texttt{T2} to
execute \texttt{\_remove()} again, thus causing the double-deletion
crash.

This runtime failure may also lead to a DoS attack, e.g., when
multiple threads keep on removing the same list entry.

\subsection{Order Violation in Memory Initialization}
\label{sec:ov_pattern}

\begin{figure}
	\centering
	\resizebox{0.48\textwidth}{!}{
		\begin{tikzpicture}
		\tikzstyle{every node}=[font=\small\ttfamily]
		\node[draw, align=left, text width=5.2cm ] (User thread) at (0, 0) {\\
			{\textcolor{darkgreen}{/* Thread T1 */}}\\
			{\tiny 1} \_user\_tselect(\textcolor{blue}{struct} file *file, ...)\{\\
			{\tiny 2} \textcolor{red!0}{00}\textcolor{blue}{struct} snd\_timer\_user *tu;\\
			{\tiny 3} \textcolor{red!0}{00}tu = file->private\_data;\\
			{\tiny 4} \textcolor{red!0}{00}......\\
			{\tiny 5} \textcolor{red!0}{00}kfree(\textcolor{magenta}{tu->queue});\\
			{\tiny 6} \textcolor{red!0}{00}\textcolor{magenta}{tu->queue} = NULL;\\
			{\tiny 7} \textcolor{red!0}{00}kfree(\textcolor{magenta}{tu->tqueue});\\
			{\tiny 8} \textcolor{red!0}{00}\textcolor{magenta}{tu->tqueue} = NULL;\\
			{\tiny 9} \textcolor{red!0}{00}{\textcolor{blue}{if}} (tu->tread) \{\\
			{\tiny 10}\textcolor{red!0}{0000}\textcolor{magenta}{tu->tqueue} = kmalloc(...);\\
			{\tiny 11}\textcolor{red!0}{0000}......\\
			{\tiny 12}\textcolor{red!0}{00}\} else \{\\
			{\tiny 13}\textcolor{red!0}{0000}\textcolor{magenta}{tu->queue} = kmalloc(...);\\
			{\tiny 14}\textcolor{red!0}{0000}......\}\\
			{\tiny 15}  \}
		};
		\node[draw, align=left,text width=6.1cm] (Attacker thread) at (6.1, 0.2) {\\
			{\textcolor{darkgreen}{/* Thread T2 */}}\\
			{\tiny 16} \_user\_read(\textcolor{blue}{struct} file *file, ...)\{ \\
			{\tiny 17} \textcolor{red!0}{00}\textcolor{blue}{struct} snd\_timer\_user *tu;\\
			{\tiny 18} \textcolor{red!0}{00}tu = file->private\_data;\\
			{\tiny 19} \textcolor{red!0}{00}......\\
			{\tiny 20} \textcolor{red!0}{00}mutex\_lock(\&tu->ioctl\_lock);\\
			{\tiny 21} \textcolor{red!0}{00}{\textcolor{blue}{if}} (tu->tread) \{\\
			{\tiny 22} \textcolor{red!0}{0000}if (copy\_to\_user(buffer, \\
			\textcolor{red!0}{000000}\textcolor{magenta}{\&tu->tqueue[qhead]},...)) \{...\}\\
			{\tiny 23} \textcolor{red!0}{00}\} else \{\\
			{\tiny 24} \textcolor{red!0}{0000}{\textcolor{blue}{if}} (copy\_to\_user(buffer, \\
			\textcolor{red!0}{000000}\textcolor{magenta}{\&tu->queue[qhead]},...))\{...\}\\
			{\tiny 25} \textcolor{red!0}{00}\}\\
			{\tiny 26} \textcolor{red!0}{00}mutex\_unlock(\&tu->ioctl\_lock);\\
			{\tiny 27} \textcolor{red!0}{00}......\\ 
			{\tiny 28} \}};
		\draw[->, dashed, red, thick] (1.65, -1.0) -- (3.5, 0.1);
		\draw[->, dashed, red, thick] (1.55, -2.0) -- (3.5, -0.9);
		\end{tikzpicture}}
	\caption{Order violation related APIs in sound/core/timer.c}
	\label{fig:timer}
\end{figure}

So far we focus only on vulnerabilities that cause noticeable system
failures, but there is another type, called \emph{Gain Information}
(GI), that may be more subtle in that it does not have obvious
symptoms. While GI seems benign, it can silently disclose sensitive
kernel data and thus lead to subsequent attacks.
In this section we study how order violation may lead to GI, as shown
in CVE-2017-1000380~\cite{CVE-2017-1000380}, which allows a local
users to read sensitive information of others.

Figure~\ref{fig:timer} shows kernel APIs \texttt{\_user\_tselect()}
and
\texttt{\_user\_read()} from the ALSA subsystem, with the same 
parameter \texttt{*file}.  The parameter has
a \texttt{snd\_timer\_user} pointer field as
\texttt{tu} (Line 3 and Line 18, respectively).   The function \texttt{\_user\_tselect()} 
re-allocates heap buffer for the queue object \texttt{tqueue}
or \texttt{queue} in \texttt{tu}, depending on the flag \texttt{tread}
(Lines 9-14). To avoid memory leaks, \texttt{\_user\_tselect()} also
frees and resets the two pointers before the new memory allocation
(Lines 5-8).

The function \texttt{\_user\_read()} copies the \texttt{tqueue}
or \texttt{queue} data from kernel space to user space by
calling \texttt{copy\_to\_user()} (Lines 22 and 24, respectively). The
source of the copy operation also depends on the flag
\texttt{tu->tread}.  To protect the data-copy  against 
unwanted interruption, \texttt{\_user\_read()}
locks \texttt{ioctl\_lock} (Line 20) until it finishes the data
transmission (Line 26).

The problem occurs if \texttt{copy\_to\_user()} is invoked right after
\texttt{\_user\_tselect()} allocates the new memory, as shown by 
the red dotted lines in Figure~\ref{fig:timer}.  Note that the kernel
API \texttt{kmalloc()} does not initialize its allocated memory, 
thus the remaining data in the newly allocated memory would be
directly copied to the buffer in the user space.  The problem is due
to an order violation where the concurrent copy operation violates the
intended order between the \texttt{kmalloc} invocation and the 
subsequent memory operation.

To trigger the leak, an attacker may construct a two-threaded program
similar to the code introduced in section~\ref{sec:av_pattern}, where
the two threads operate on the same file pointer by repeatedly
invoking \texttt{\_user\_tselect()} and
\texttt{\_user\_read()}, respectively.  Once the aforementioned 
interleaving occurs, uninitialized memory content will be leaked to the
user space.

\subsection{Discussion}

Concurrent vulnerabilities are common in Linux kernel code. The three
case studies presented in this section represent three different
concurrency bug types.  They also lead to different types of system
failures such as hung, program crash, and information leak.
Often times, these system failures are correlated with each other,
which means it is not always possible to classify a concurrency
vulnerability into any one type (see
Section~\ref{sec:Vulnerability-result} for details).

The erroneous thread interleavings that we present are not necessarily
the only possible ways of exploiting these vulnerabilities.  There may
be many other possible interleavings.  In fact, concurrency often
leads to a large number of possible interleavings, some of which may
be exploited; this is also one of the reasons why concurrency vulnerability
detection and mitigation may be difficult.

\section{Characteristics of  Vulnerabilities}
\label{sec:Vulnerability-result}

In this section, we present the statistics of the 101 concurrency
vulnerabilities used in our study, including their bug types,
vulnerability types, adversarial types, as well as the numbers of
threads and shared variables.  In the next section, we provide similar
statistics for the corresponding repair strategies.

\subsection{Distribution of the Samples}

\begin{figure}
	\centering
	\includegraphics[width=.4\textwidth]{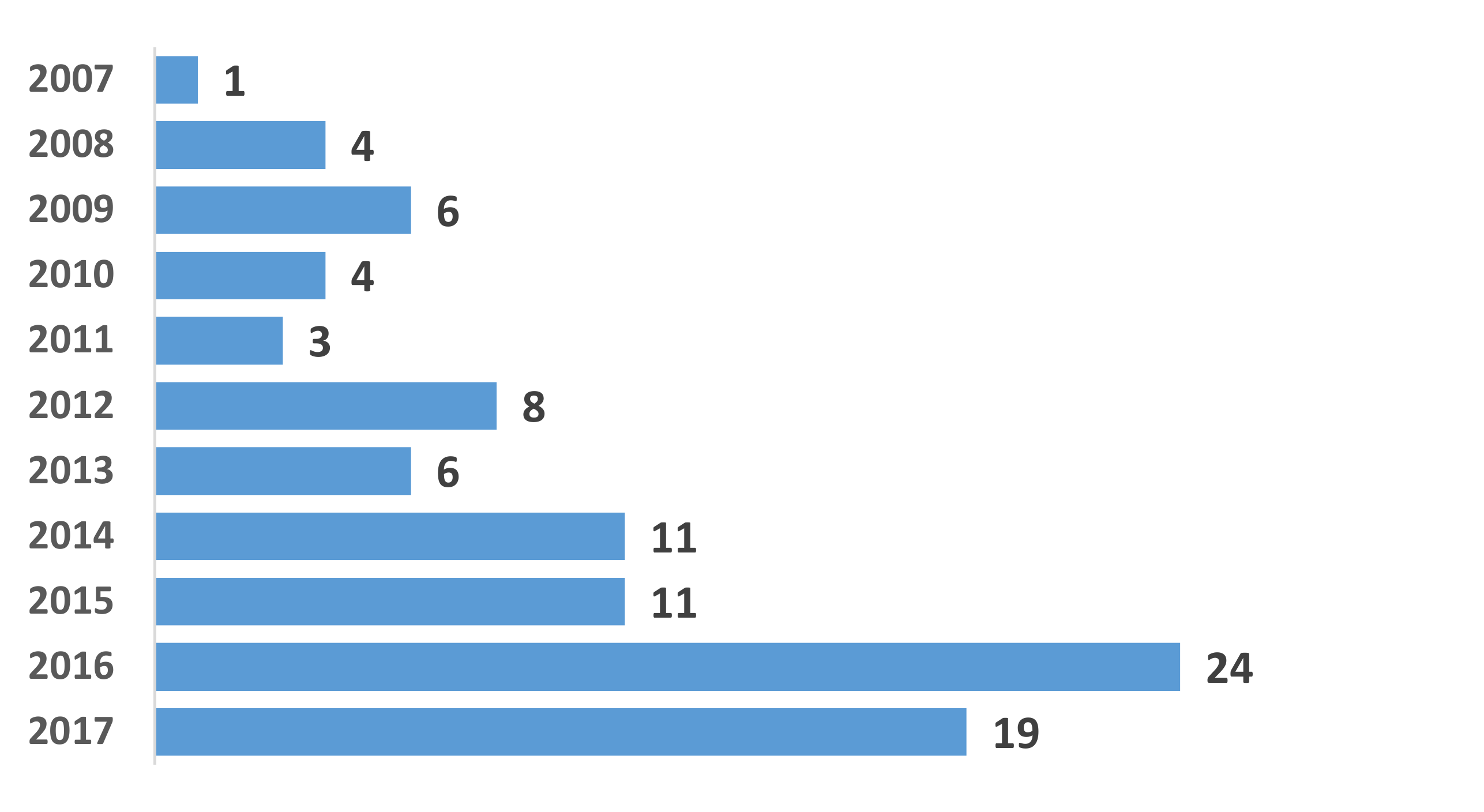}
	\caption{Number of concurrency vulnerabilities per year.}
	\label{fig:convul-years}
\end{figure}

Our vulnerability samples are from the past ten
years. Figure~\ref{fig:convul-years} shows the number in each year. We omit the entries of the year 2018
because our data gathering was conducted in the summer of 2018.  The
result shows that, overall, the number of concurrency vulnerabilities
is increasing.

\begin{figure}
	\centering
	\includegraphics[width=.475\textwidth]{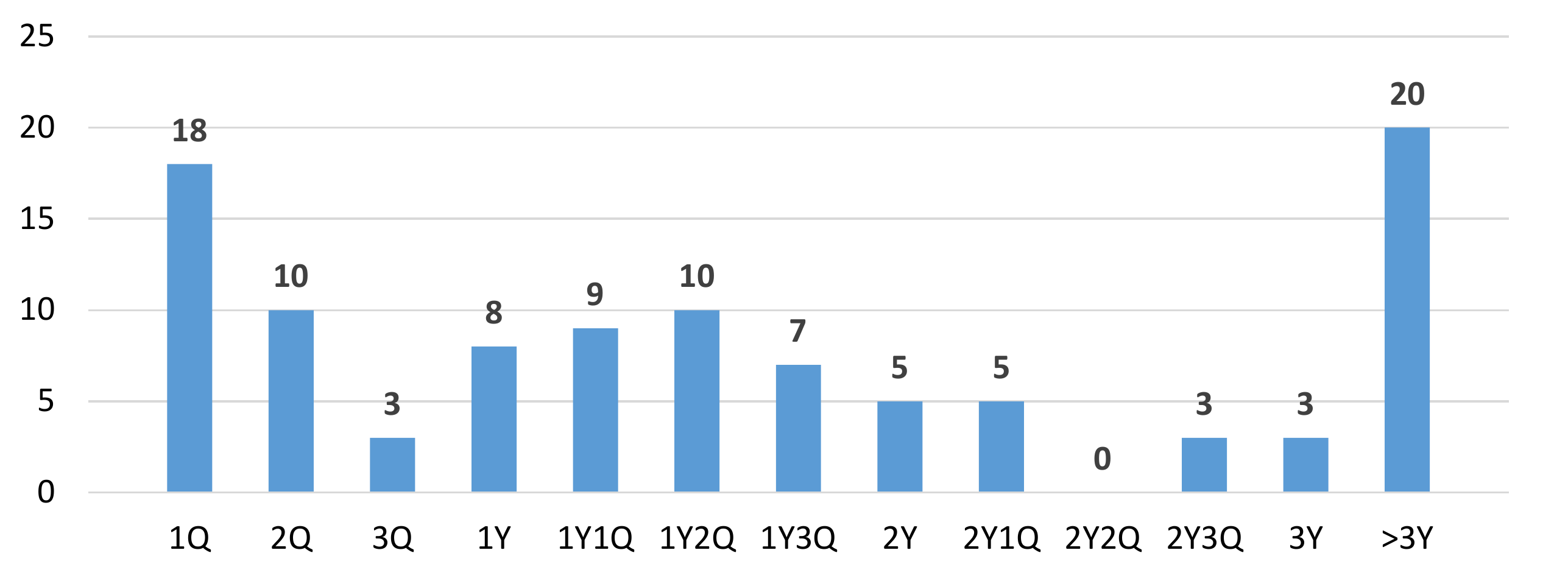}
	\caption{Distribution of time taken to patch a vulnerability.}
	\label{fig:duration}
\end{figure}

To understand how long it takes for developers to patch a vulnerability,
we compute the duration between the \emph{publishing date} and
the \emph{last modified} date.  The longer the duration, the more time
is needed by developers to understand and patch the vulnerability.
Figure~\ref{fig:duration} shows the distribution of 101 vulnerabilities over 
different patch durations from 1 quarter up to 3 years. The distribution
shows a high distribution on both sides, indicating
a vulnerability will either be relatively easy to patch, i.e., in a short period of time, or 
so difficult that it cannot be patched within a coupe of years.

Table \ref{tab:cve-modify} shows how many patches are needed before a
vulnerability is eliminated.  Only 19 of the 101 vulnerabilities (or
18\%) were eliminated by the first patch.  In contrast, 37 of them (or
36\%) required more than five patches.  This is also alarming because
it shows how difficult it is to construct a patch.  Here, more patches
generally mean that the vulnerability is more complex.  Since
reasoning about thread interaction is challenging, developers often
need to construct the patch in a trial-and-error fashion.

\begin{table}[htbp]
	\centering
	\caption{Number of patches needed to fix a vulnerability.}
	\label{tab:cve-modify}
	\scalebox{.75}{
		\begin{tabular}{|c|c|r|}
			\hline
			Number of  Patches Needed & Number of Vulnerabilities & Percentage \\ \hline\hline
			1 & 19 & 18.81\% \\ \hline
			2 & 9 & 8.91\%  \\ \hline
			3 & 16 & 15.84\%  \\ \hline
			4 & 12 & 11.88\%  \\ \hline
			5 & 8 & 7.92\%  \\ \hline
			$>$5 & 37 & 36.63\%  \\ \hline
		\end{tabular}
	}
\end{table}

\subsection{Distribution of Bug Types}

\begin{figure}
	\begin{minipage}{.45\linewidth}	
		\vspace{-2.5ex}
		\includegraphics[width=\textwidth]{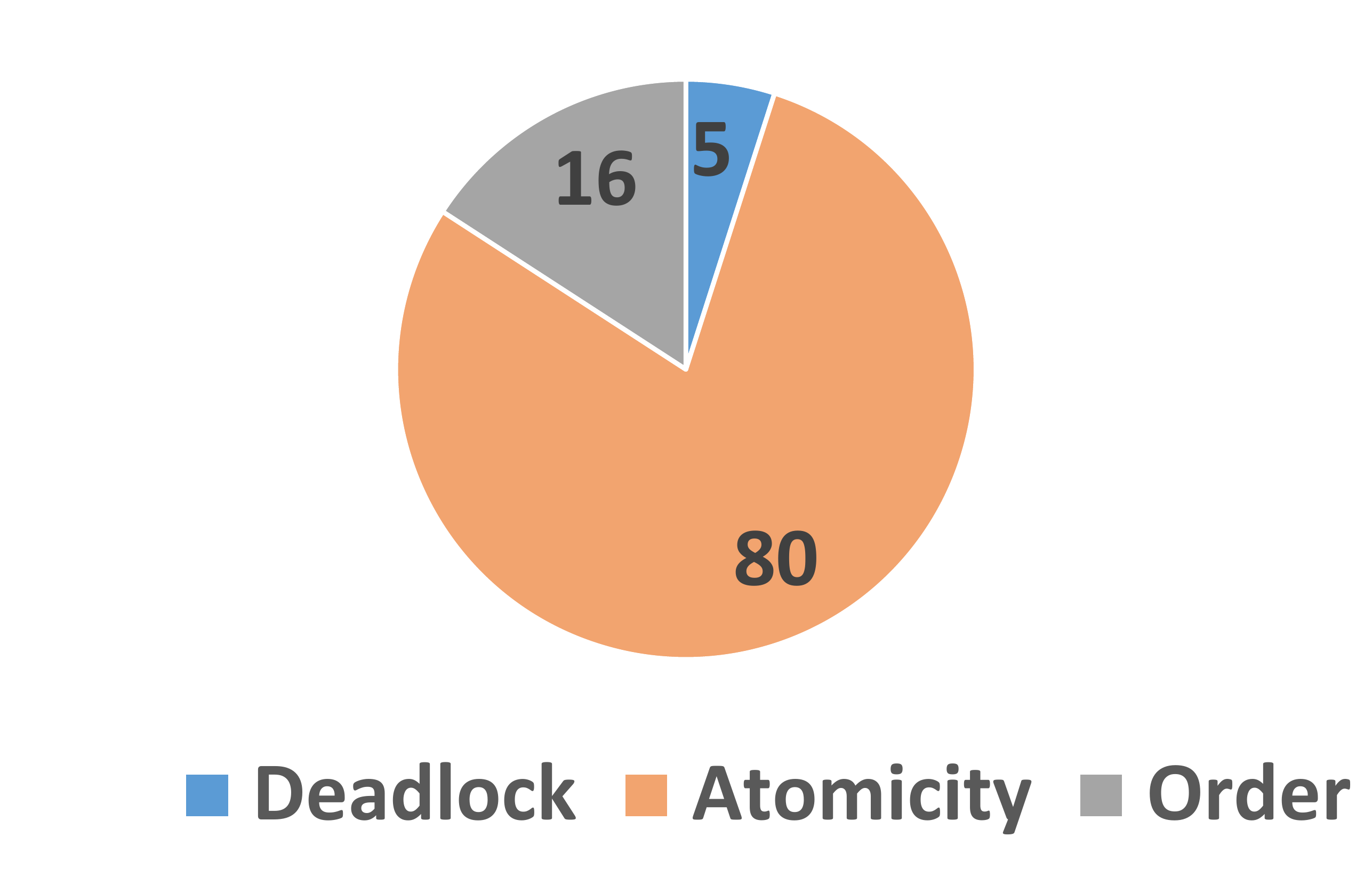}
	\end{minipage}
	\begin{minipage}{.45\linewidth}	
		\includegraphics[width=\textwidth]{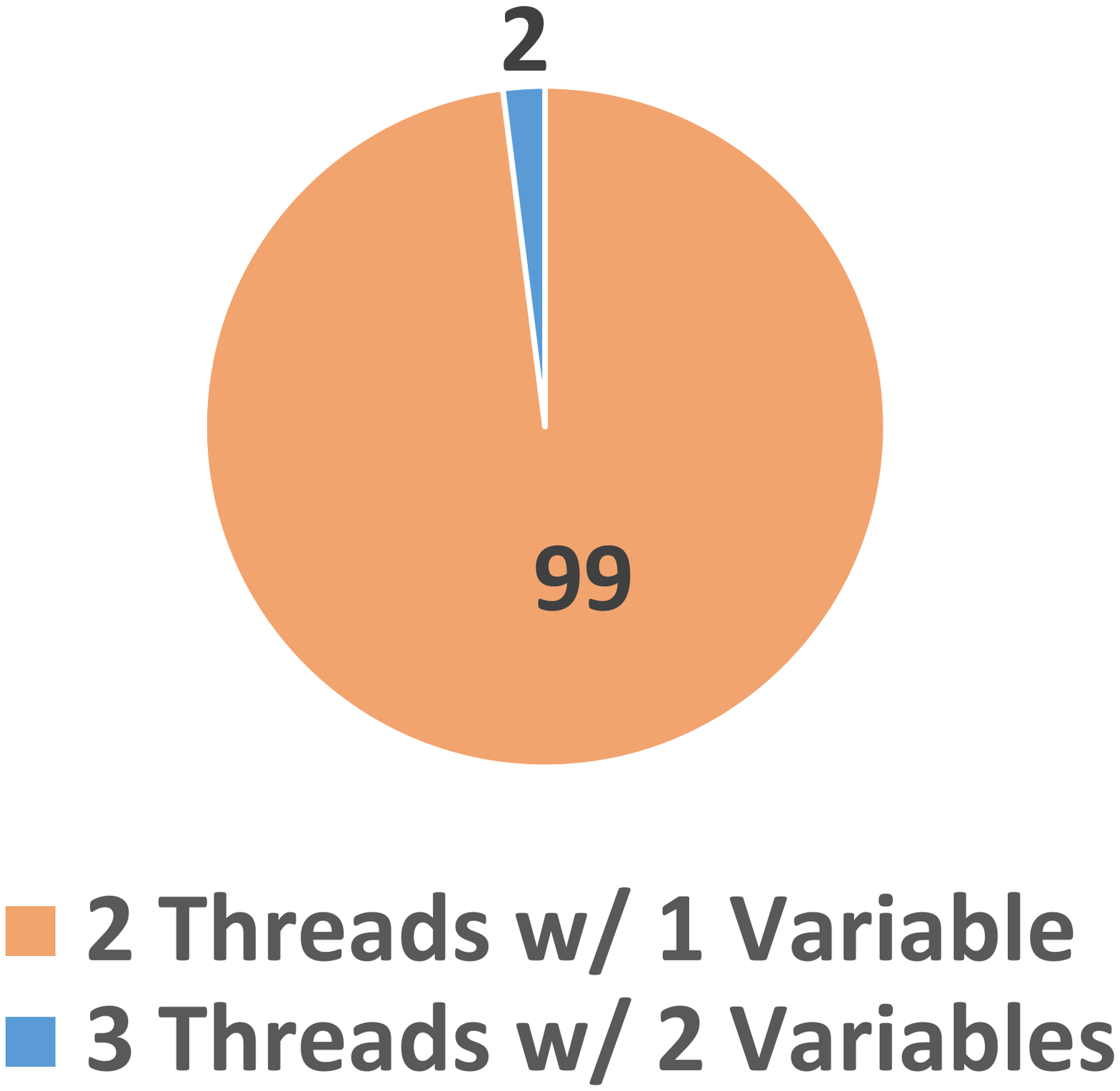}
	\end{minipage}
	\caption{Distribution of concurrency bug types (left) and the number of involved threads and shared variables (right).}
	\label{fig:dist}
\end{figure}

Next, we present characteristics of the concurrency bugs
associated with the 101 vulnerabilities.  Figure~\ref{fig:dist} shows
the distribution of bug types on the left-hand size, and the number of
threads and variables on the right-hand side. 
The result shows that the vast majority (95\%) are non-deadlock bugs,
whereas deadlocks account for only 5\%.  Furthermore, among the
non-deadlock bugs, there are five times more atomicity violations
than order violations.

As for the number of threads and variables involved in each
vulnerability, we find that only 2 of the 101 samples clearly involve
more than two threads and one variable. The rare occurrence
of more than two threads and one variable suggests that we,
as a community, should focus more on these cases during
vulnerability detection and repair.

\subsection{Distribution of Vulnerability Types}

Next, we present statistics on the adversarial types and vulnerability
types.  For adversarial types, we want to know whether the attacker
needs to be a local user, e.g., a user who can execute non-privileged
programs on the victim's computer, or a remote user who has to access
the victim's server.  Table \ref{tab:adv-type-distribution} shows the
distribution of these two types of adversaries.  The result shows that
only 10\% of concurrency vulnerabilities are exploitable by a remote
user.

\begin{table}[htbp]
	\centering
	\caption{Distribution of the adversary types.}
	\label{tab:adv-type-distribution}
	\scalebox{.75}{
		\begin{tabular}{|p{.45\linewidth}|c|r|}
			\hline
			Adversarial Type & Number of  Vulnerabilities & Percentage \\ \hline\hline
			Local Attacker  & 90 & 89.11\% \\ \hline
			Remote Attacker & 11 & 10.89\% \\ \hline
		\end{tabular}
	}
\end{table}

Table \ref{tab:vul_dist} shows the distribution of vulnerability
types. Among the 101 samples, 72\% of them may lead to Denial of Service
(DoS), 69\% of them lead to Memory Corruption (MC), and 30\% of them
lead to Gain Privilege (GP).  Each of the other three types accounts
for less than 10\%.
Among these six vulnerability types, GP (30\%) is particularly
alarming, because the operating system cannot afford to have such a
high risk of losing privileges to attackers.  
While GI (6\%) and BS (5\%) are also rare, they should not be neglected
because they are often the first steps of a sequence of deadly
attacks.

\begin{table}[!htbp]
	\centering
	\caption{Distribution of the vulnerability types.}
	\label{tab:vul_dist}
	\scalebox{.75}{
		\begin{tabular}{|l|c|c|r|}
			\hline
			{Vulnerability Type} & {Abbreviation}& {Num. Vulnerabilities} & {Percentage}\\
			\hline\hline
			Denial of Service & DoS & 73 & 72.28\%\\
			\hline
			Memory Corruption & MC & 61 & 60.40\%\\
			\hline
			Gain Privilege & GP & 30 & 29.70\%\\
			\hline
			Gain Information & GI & 6 & 5.94\%\\
			\hline
			Bypass Something & BS &5 & 4.95\%\\
			\hline
			Execute Code & EC & 2 & 1.98\%\\
			\hline
		\end{tabular}
	}
\end{table}

\begin{figure}
	\centering
	\includegraphics[width=.25\textwidth]{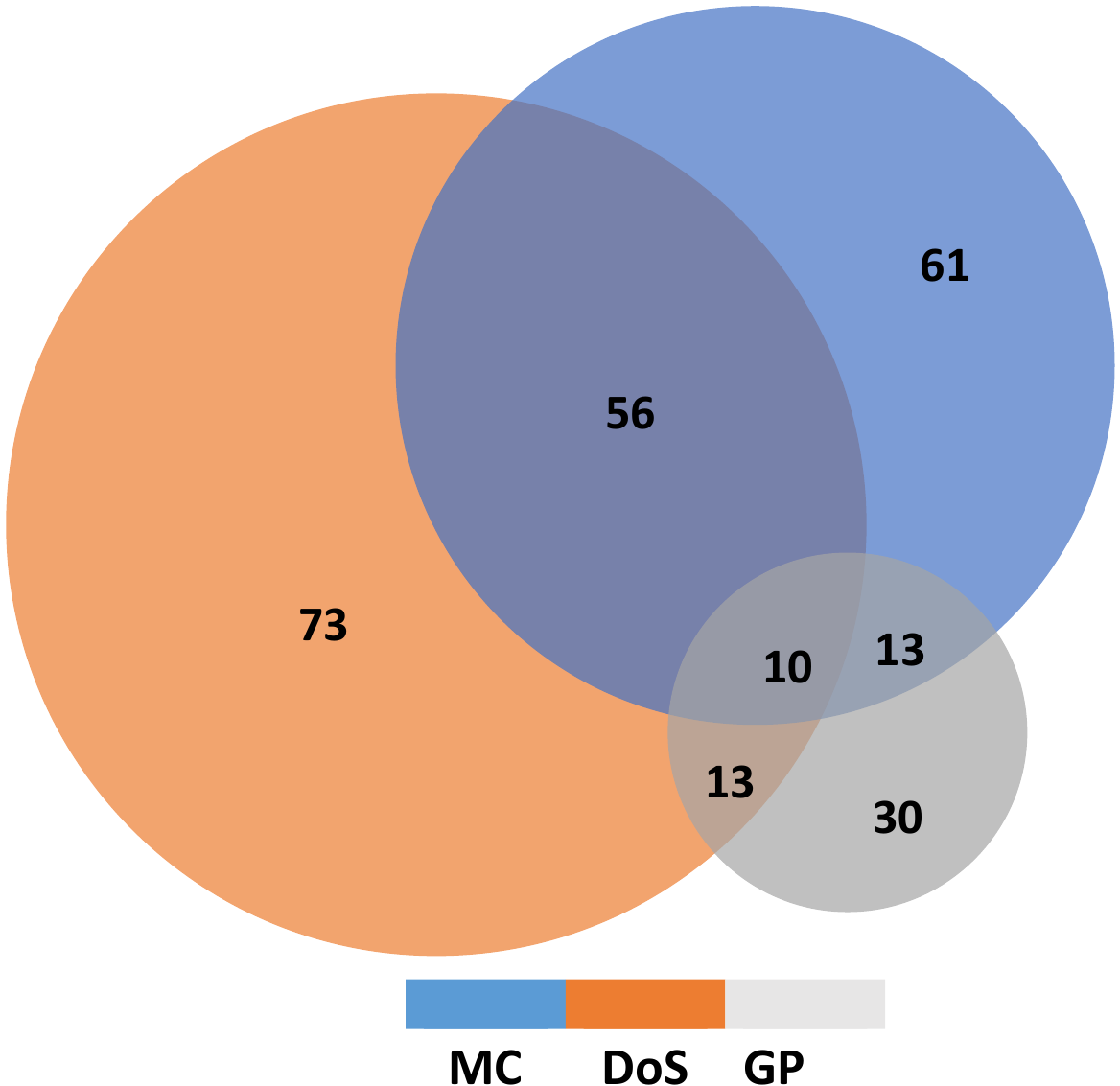}
	\caption{Overlapping vulnerability types: MC--DoS--GP. }
	\label{fig:vul-venn}
\end{figure}

Since the vulnerability types are not mutually exclusive, we
want to know how much they overlap.  Figure~\ref{fig:vul-venn}
shows the Venn diagram of the three largest groups: DoS, MC, and GP,
together with the number of vulnerabilities in each group.  Among the
101 samples, 56 may lead to both DoS and MC, 23 may lead to both DoS
and GP, and 23 may lead to both MC and GP.  Interestingly, there are
10 vulnerabilities that may lead to all three types.  A closer look at
these 10 vulnerabilities show that MC could potentially grant the
attacker more privilege in the system and then cause the system to
hang.

Table \ref{tab:vul-type-vs-manifestation} shows the manifestations of
each vulnerability type, i.e., the number of variables and threads
involved in triggering the vulnerability.  The result shows that
nearly all of them involve two threads and one shared variable.  For
the vulnerabilities that involve more than two threads and more than
one variable, a closer look shows that they can only lead to DoS and
GP, perhaps because they sometimes require more complex thread
interactions.

\begin{table}[htbp]
	\centering
	\caption{Manifestations of the vulnerability types.}
	\label{tab:vul-type-vs-manifestation}
	\scalebox{.75}{
		\begin{tabular}{|l|r|r|r|r|}
			\hline
			Vulnerability Type & \multicolumn{2}{c|}{Number of variables} &
			\multicolumn{2}{c|}{Number of threads} \\\cline{2-5}
			& one variable & $>1$ variable & two threads & $>2$ threads \\ 
			\hline\hline
			Denial of Service  & 98.63\%  &1.37\%  & 98.63\%  &1.37\%   \\ \hline
			Memory Corruption  & 100.00\% &   0\%  & 100.00\% &   0\%   \\ \hline
			Gain Privilege     & 93.33\%  &6.67\%  & 93.33\%  &6.67\%   \\ \hline
			Gain Information   & 100.00\% &   0\%  & 100.00\% &   0\%   \\ \hline
			Bypass Something   & 100.00\% &   0\%  & 100.00\% &   0\%   \\ \hline
			Execute Code       & 100.00\% &   0\%  & 100.00\% &   0\%   \\ \hline
		\end{tabular}
	}
\end{table}

Table~\ref{tab:vul-type-vs-bug-type} shows the bug type distribution
within each vulnerability type.  That is, for each vulnerability type,
it shows how many of them are caused by deadlock, atomicity
violations, and order violations. 
The main observation is that atomicity violation may lead to all kinds
of vulnerabilities, and therefore should be the focus of future
research on concurrency vulnerability detection, analysis, and repair.

\begin{table}[htbp]
	\centering
	\caption{Bug type distribution in vulnerability types.}
	\label{tab:vul-type-vs-bug-type}
	\scalebox{.75}{
		\begin{tabular}{|l|r|r|r|}			\hline
			Vulnerability Type & Atomicity Violation & Order Violation & Deadlock \\ \hline\hline
			Denial of Service & 75.34\% & 6.85\% & 6.85\%\\ \hline
			Memory Corruption & 75.41\% & 22.95\% & 1.64\% \\ \hline
			Gain Privilege & 93.33\% & 6.67\% & 0.00\% \\ \hline
			Bypass Something & 83.33\% & 16.67\% & 0.00\% \\ \hline
			Gain Information & 100.00\% & 0.00\% & 0.00\% \\ \hline
			Execute Code & 100.00\% & 0.00\% & 0.00\% \\ \hline
		\end{tabular}
	}
\end{table}

Only one case with Denial of Service and Memory Corruption vulnerability types has deadlock bug type~\cite{CVE-2018-1000004}. In that case, deadlock leads to Denial of Service or it may have use-after-free of Memory Corruption vulnerability.  


\section{Study of Repair Strategies}
\label{sec:repair-type}

In this section, we study the patches adopted by developers for
eliminating the 101 vulnerabilities. After carefully following the
online descriptions, discussions and the related code, we identify
four repair strategies, named \emph{condition check}, \emph{code
	switch}, \emph{lock strategy} and \emph{design change}, respectively.
Table~\ref{tab:repair-strategy-distribution} shows the overall
distribution of these four strategies.  Among them, \emph{condition
	check} is the most frequently used, accounting for almost half of all
patches, whereas \emph{code switch} is the least frequently used,
accounting for less than 10\% of the patches.

\begin{table}[htbp]
	\centering
	\caption{Distribution of the repair strategies.}
	\label{tab:repair-strategy-distribution}
	\scalebox{.75}{
		\begin{tabular}{|p{.45\linewidth}|c|r|}
			\hline
			Patching Strategy & Number of Vulnerabilities  & Percentage \\ \hline\hline
			Condition Check & 48 & 47.52\% \\ \hline
			Code Switch & 9 & 8.91\% \\ \hline
			Lock Strategy & 29 & 28.71\% \\ \hline
			Design Change & 39 & 38.61\% \\ \hline
		\end{tabular}
	}
\end{table}

In the remainder of this section, we use examples to illustrate each
individual repair strategy.

\subsection{Condition Check}

Condition check is an intuitive way of changing the intra-thread
control flow as well as the inter-thread interleaving, thus avoiding
the unsafe program states.  For example, in the aforementioned
CVE-2016-5195 (Dirty COW), developers adopted a patch constructed
by adding a condition check. 

Recall that in Dirty COW, the problem comes from the unexpected
removal of a specific \emph{Get User Pages Flags}
(or \texttt{gup\_flags}) during the handling of page faults, which is
then exploited to grant the write access to privileged files.  

One way to address the problem is introducing a new internal flag,
named \texttt{FOLL\_COW}, for \texttt{gup\_flags}; the new flag
indicates the completion of the Copy-On-Write process.

Figure~\ref{ls:cond} shows the difference between the original and
modified versions of a related code snippet, where the condition check
is inserted to validate the page table entries (\emph{pte})
checking: \emph{pte} can only be writable after going through a COW
cycle.  By adding this condition check, the code will be able to
properly handle the case where COW-ed page is purged by a malicious
thread.

\begin{figure}
	\centering
	\begin{minipage}[b]{.9\linewidth}	
		\begin{lstlisting}[language=diff, numbers=none]
		@@ -60,6 +60,16 @@ static int follow_pfn_pte(
		#define FOLL_FORCE 0x10 
		+ #define FOLL_COW   0x4000
		
		+ static inline bool can_follow_write_pte(...){
		+     return pte_write(pte) || 
		+            ((flags & FOLL_FORCE) && 
		+             (flags & FOLL_COW) && 
		+              pte_dirty(pte)); 
		+ }
		static struct page *follow_page_pte(...){
		if ((flags & FOLL_NUMA) && pte_protnone(pte))
		goto no_page;
		-     if ((flags & FOLL_WRITE) && !pte_write(pte)) {
		+     if ((flags & FOLL_WRITE) && 
		+           !can_follow_write_pte(pte, flags)) {
		pte_unmap_unlock(ptep, ptl);
		return NULL;
		}
		\end{lstlisting}
	\end{minipage}
	\vspace{-1ex}
	\caption{Patch for Dirty COW in mm/gup.c.}
	\label{ls:cond}
\end{figure}

\subsection{Code Switch}

Another strategy for eliminating concurrency vulnerabilities is
switching the order of certain code statements.  Consider the patch of
CVE-2016-9794 as an example, which demonstrates this strategy.

Here, a function named \texttt{kill\_fasync()} is provided by the
Digital Audio (PCM) library in Linux kernel.  As shown in
Figure~\ref{ls:switch}, the vulnerable code invokes this function from
outside of the stream lock protected region, which may lead to a race
condition.  

The quick workaround adopted by Linux kernel developers is to move the
call to \texttt{kill\_fasync()} inside the region protected by the
stream lock.  Although this \emph{one-liner} may enlarge the scope of
the stream lock protected region, and thus in theory lead to
performance degradation, in practice, it actually has little impact on
the system's performance. 

\begin{figure}
	\centering
	\begin{minipage}{.9\linewidth}	
		\begin{lstlisting}[language=diff, numbers=none]
		@@ -1886,8 +1886,8 @@ void snd_pcm_period_elapsed(
		snd_timer_interrupt(substream->timer, 1);
		#endif
		_end:
		-     snd_pcm_stream_unlock_irqrestore(substream, flags);
		kill_fasync(&runtime->fasync, SIGIO, POLL_IN);
		+     snd_pcm_stream_unlock_irqrestore(substream, flags);
		}
		EXPORT_SYMBOL(snd_pcm_period_elapsed);
		\end{lstlisting}
	\end{minipage}
	\vspace{-1ex}
	\caption{Patch for \emph{use-after-free} in sound subsystem.}
	\label{ls:switch}
\end{figure}

\subsection{Lock Strategy}

The third strategy for eliminating concurrency
vulnerabilities is to change the way locks are used.  For example, a
new lock may be inserted to wrap up an entire critical section, to
avoid it being interrupted by other threads.  However, new locks must
be added with caution since they often lead to performance
degradation.  This is especially important for Linux kernel developers
who care about performance. 

Figure~\ref{ls:lock} shows a patch constructed for CVE-2017-10661,
which has been discussed in Section~\ref{sec:av_pattern}.  Here, a new
lock-unlock pair is added to protect the critical region, to ensure
that the concurrent operations on \texttt{ctx} inside \texttt{setup()}
and \texttt{remove()} are always executed atomically.

\begin{figure}	
	\centering
	\resizebox{1.02\linewidth}{!}{
		\begin{tikzpicture}
		\tikzstyle{every node}=[font=\footnotesize\ttfamily]
		\node[draw, align=left,text width=4.8cm] (timerfd) at (0,1.7) {\\
			{\tiny 1} {\color{black}{struct}} timerfd\_ctx \{\\
			{\tiny 2} \textcolor{red!0}{00} ......\\
			{\tiny 3} \textcolor{red!0}{00} {\color{black}{struct}} list\_head clist;\\
			{\tiny +} \textcolor{red!0}{00} \textcolor{darkblue}{spinlock\_t cancel\_lk;}\\ 
			{\tiny 4} \textcolor{red!0}{00} {\color{black}{bool}} might\_cancel; \}\\
		};
		\node[draw, align=left,text width=4.8cm] (usr thread) at (0,-1.0) {\\
			{\tiny 5} \_remove (timerdf\_ctx *ctx) \{\\
			{\tiny +} \textcolor{red!0}{00}\textcolor{darkblue}{spin\_lock(\&ctx->cancel\_lk);}\\ 
			{\tiny 6} \textcolor{red!0}{00}{\color{black}{if}}(ctx->might\_cancel)\{\\
			{\tiny 7} \textcolor{red!0}{000}ctx->might\_cancel = false;\\
			{\tiny 8} \textcolor{red!0}{000}spin\_lock(\&cancel\_lock);\\
			{\tiny 9} \textcolor{red!0}{000}list\_del\_rcu(\&ctx->clist);\\
			{\tiny 10} \textcolor{red!0}{00}spin\_unlock(\&cancel\_lock);\\
			{\tiny 11} \textcolor{red!0}{0}\}\\
			{\tiny +}  \textcolor{red!0}{00}\textcolor{darkblue}{spin\_unlock(\&ctx->cancel\_lk);}\\ 
		};
		\node[draw, align=left,text width=5.2cm] (attacker thread) at (5.3, -0.0) {\\
			{\tiny 12} \_setup (timerfd\_ctx *ctx, ...)\{\\
			{\tiny 13} \textcolor{red!0}{00}......\\ [0.2em]
			{\tiny  +}  \textcolor{red!0}{000}\textcolor{darkblue}{spin\_lock(\&ctx->cancel\_lk);}\\ 
			{\tiny 14} \textcolor{red!0}{00}{\color{black}{if}}(!ctx->might\_cancel)\{\\
			{\tiny 15} \textcolor{red!0}{0000}ctx->might\_cancel = true;\\
			{\tiny 16} \textcolor{red!0}{0}\textcolor{red!0}{000}spin\_lock(\&cancel\_lk);\\
			{\tiny 17} \textcolor{red!0}{0000}list\_add\_rcu(\&ctx->clist,\\
			\textcolor{red!0}{000000}\&cancel\_list);\\
			{\tiny 18} \textcolor{red!0}{0}\textcolor{red!0}{000}spin\_unlock(\&cancel\_lk);\\
			{\tiny 19} \textcolor{red!0}{00}\}{\color{black}{else if}}(ctx->might\_cancel)\{\\
			{\tiny 20} \textcolor{red!0}{0000}\_remove({ctx});\\
			{\tiny 21} \textcolor{red!0}{00}\}\\
			{\tiny 22} \textcolor{red!0}{00}......\\[0.2em]
			{\tiny  +} \textcolor{red!0}{000}\textcolor{darkblue}{spin\_unlock(\&ctx->cancel\_lock);}\\ 
			{\tiny 23} \}\\
		};
		\end{tikzpicture}
	}
	\caption{Patch for race condition in \emph{timerfd} subsystem.}
	\label{ls:lock}
\end{figure}

\subsection{Design Change}

The last repair strategy that developers used to eliminate concurrency
vulnerabilities is design change.  That is, developers choose to solve
the security issues by refactoring the program, which in turn changes
the thread synchronization patterns as well as the related data
structures.  Although the complexity of design change may vary, e.g.,
ranging from simply removing a few unnecessary variables from the
shared data structure to reimplementing the entire functionality, in
general, it is significantly more complex than the other three
strategies.  Therefore, it is also the most difficult to automate.

For example, in CVE-2016-2545~\cite{CVE-2016-2545}, recall that the
problem is due to \texttt{snd\_timer\_interrupt()} from Advanced Linux
Sound Architecture (ALSA) subsystem, which does not properly maintain
a couple of linked lists.  Therefore, it allows a local user to
trigger a race condition and then a system crash.  In this case, the
linked lists may be removed by another function
named \texttt{snd\_timer\_stop()} while it is being removed by the
function \texttt{snd\_timer\_interrupt()}, thus resulting in a
double-free.

The patch adopted by Linux developers added a brand-new function
named \texttt{list\_del\_init()}, which is used to remove the list.
Then, all other functions delegate the deletion operation to this new
function.  They can avoid the double-free error by always
re-initializing the list entry after deletion.

\subsection{Distribution of Repair Strategies}

\begin{figure}
	\centering
	\includegraphics[width=.45\textwidth]{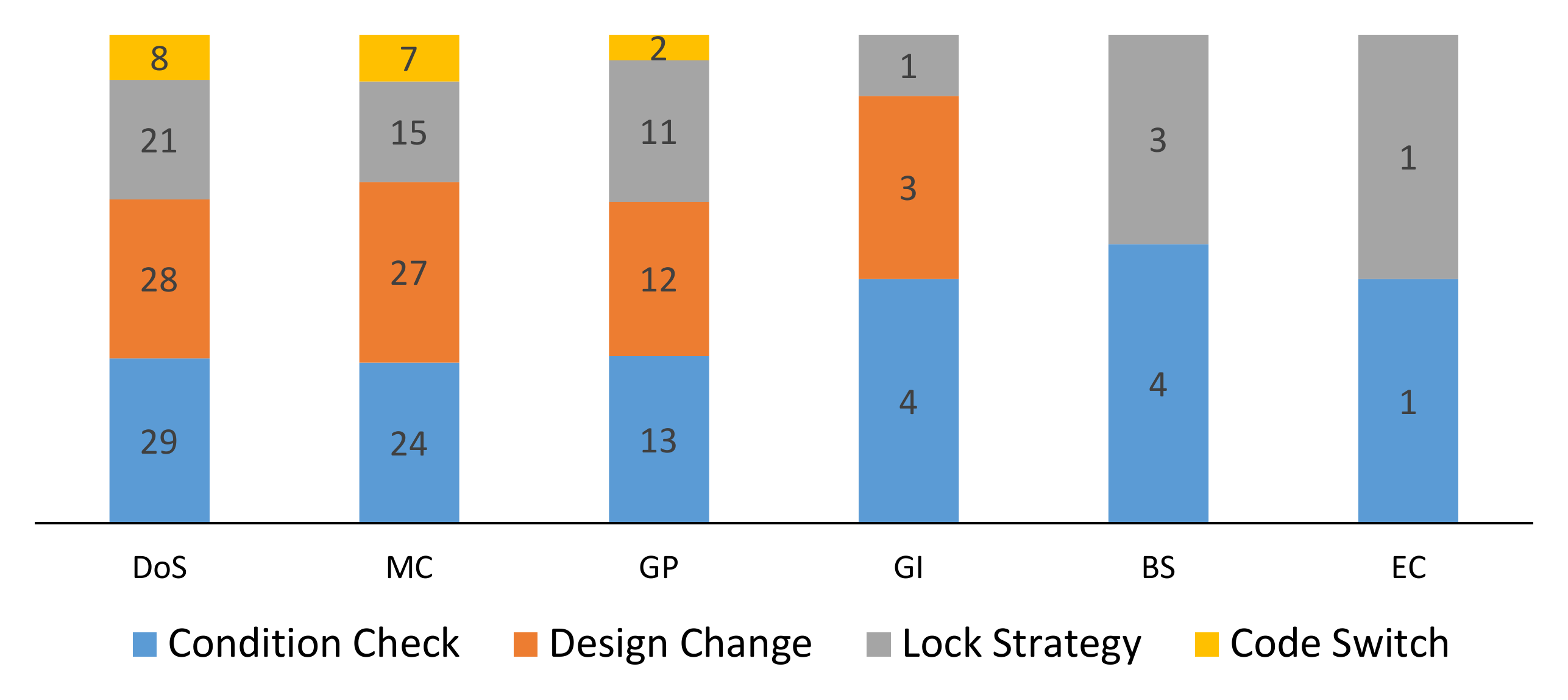}
	\caption{Repair strategy distribution in vulnerabilities.}
	\label{fig:repairDist}
\end{figure}

We analyze the patches adopted by Linux developers for the 101
concurrency vulnerabilities, and count the number of times each of the
four repair strategies are used for each vulnerability type.  The
result is shown in Figure~\ref{fig:repairDist}.  That is, for every
vulnerability type, we show the number of times that each repair
strategy is used to construct the patches.

The result shows that, overall, \emph{condition check} is the most
frequently used repair strategy; \emph{lock strategy} is also used to
construct patches for all vulnerability types.  However, since adding
locks may lead to severe performance degradation, it is not as
frequently used by kernel developers as they are used by application
developers~\cite{aldrich2003comprehensive,bogda1999removing,ruf2000effective}. This
is perhaps because kernel developers are more cautious about the
patch's performance impact.
In contrast, \emph{design change} has been used for DoS, MC, GP and GI
only.  Finally, \emph{code switch} has been used for DoS, MS, and GP
only.  Although it may appear to be simple, \emph{code switch} can
become tricky and thus may not be applicable in complex situations
where it is difficult to determine whether swapping the order of
program statements may fix certain issues.

Overall, our result shows that patching concurrency vulnerabilities
remain a challenging task: the duration between a vulnerability's
publishing date and the last-modification date is long. Thus, we need to devote more effort in developing automated tools and
techniques, to help speed up the vulnerability detection,
analysis, and repair.

\section{Related Work}
\label{sec:related}

There is a large body of work on studying various aspects of
concurrency bugs in software~\cite{Lu2008, Fu2018, Stender2008, CaiC12, CaiWC14, GuoCY17, ZhouSLCL17, CaiCZ17,
	Paleari2008, Wang2017, Warszawski2017, Watson2007, Wei2005, Yang2012, HuangZhang2016}. 
As we mentioned earlier, Lu et al.~\cite{Lu2008} conducted a
comprehensive study of real-world concurrency bug characteristic in
four open-source applications: Mozilla, Apache, MySQL and OpenOffice,
and their bug samples were from the years prior to 2007.  Fu et
el.~\cite{Fu2018} surveyed more than 100 research papers, together with
10 applications, based on which they analyzed strategies used for
automated bug detection, replay, and avoidance. However, none of these
prior works focus on vulnerabilities caused by concurrency bugs.

In different application domains, Stender et al.~\cite{Stender2008}
and Paleari et al.~\cite{Paleari2008} studied concurrency
vulnerabilities, however, their focus was on either on web
applications or distributed database systems, not on 
vulnerabilities in Linux kernel code. Yang et al.~\cite{Yang2012}
conducted a summary and categorization of a limited number of
concurrency attacks.  Warszawski et al.~\cite{Warszawski2017} also
focused on concurrency vulnerabilities in database systems,
which may be exploited by APIs accessible in programs that use the
database.  However, their focus was on detecting these vulnerabilities
using a trace based static analysis as opposed to empirical studies.

Wang et al.~\cite{Wang2017} investigated conditions under which
double-fetch may become vulnerabilities in Linux kernels. They also
provided a technique for detecting such double-fetch
vulnerabilities.  Wei et al.~\cite{Wei2005} and Watson et
al.~\cite{Watson2007} both studied the TOCTTOU vulnerabilities
in Linux, which represent a small subset of the possible concurrency
attacks, and their focus was on identifying
vulnerabilities that may lead to privilege escalations or
bypassing security checks. In contrast to these prior works, which
touched upon only certain aspects of concurrency vulnerabilities, we
conducted the first compensative study of all three bug types, seven
vulnerability types, and 101 real vulnerabilities from Linux kernel
code.

There is also a large body of work on developing techniques for
detecting and mitigating concurrency bugs~\cite{Kasikci2012, Liu2018,
  Zhao2018, Cowan2001}.  In particular, Kasikci et
al.~\cite{Kasikci2012} studied the difference between data-races and
data-race bugs, and developed a tool for identifying harmful
data-races.  Heuristic techniques have also been applied to
concurrency bug detection~\cite{Liu2018}, which employs both static
analysis and fuzzing testing.  Zhao et al.~\cite{Zhao2018} developed a
tool named \emph{OWL}, which relies on the implicit impact of
concurrency vulnerabilities to detect them.  Cowan et
al.~\cite{Cowan2001} developed a tool named \emph{RaceGuard} to
prevent temporal file injections.  However, they were not
comprehensive studies on concurrency related security vulnerabilities.


\section{Conclusions}
\label{sec:conclusion}

We have presented a comprehensive study of 101 recent concurrency
vulnerabilities in Linux kernels. All of these vulnerabilities were
reported to the CVE database in the past ten years; they were also
confirmed, analyzed, and eventually patched by Linux kernel
developers.  We illustrated the individual concurrency bug types,
vulnerability types, and patching strategies using real-world
examples.  We also presented the main characteristics of these
vulnerabilities, including their manifestations, such as the number of
variables and threads involved, distribution of bug types, adversarial
types and vulnerability types, distribution of repair strategies, and
number of months taken for developers to fix the vulnerability.  With
the study, we hope to shed light on the severity of concurrency
vulnerabilities and potential deficiencies of existing tools, and
point out some future research directions.

\clearpage
\newpage

\bibliographystyle{plain}
\bibliography{convul}

\begin{thebibliography}{10}

\bibitem{aldrich2003comprehensive}
Jonathan Aldrich, Emin~G{\"u}n Sirer, Craig Chambers, and Susan~J Eggers.
\newblock Comprehensive synchronization elimination for {Java}.
\newblock {\em Science of Computer Programming}, 47(2-3):91--120, 2003.

\bibitem{bogda1999removing}
Jeff Bogda and Urs H{\"o}lzle.
\newblock Removing unnecessary synchronization in {Java}.
\newblock In {\em ACM SIGPLAN Notices}, volume~34, pages 35--46. ACM, 1999.

\bibitem{CaiCZ17}
Yan Cai, Lingwei Cao, and Jing Zhao.
\newblock Adaptively generating high quality fixes for atomicity violations.
\newblock In {\em Proceedings of the 2017 11th Joint Meeting on Foundations of
  Software Engineering, {ESEC/FSE} 2017, Paderborn, Germany, September 4-8,
  2017}, pages 303--314, 2017.

\bibitem{CaiC12}
Yan Cai and W.~K. Chan.
\newblock {MagicFuzzer}: Scalable deadlock detection for large-scale
  applications.
\newblock In {\em 34th International Conference on Software Engineering, {ICSE}
  2012, June 2-9, 2012, Zurich, Switzerland}, pages 606--616, 2012.

\bibitem{CaiWC14}
Yan Cai, Shangru Wu, and W.~K. Chan.
\newblock Conlock: a constraint-based approach to dynamic checking on deadlocks
  in multithreaded programs.
\newblock In {\em 36th International Conference on Software Engineering, {ICSE}
  '14, Hyderabad, India - May 31 - June 07, 2014}, pages 491--502, 2014.

\bibitem{ChouYCHE01}
Andy Chou, Junfeng Yang, Benjamin Chelf, Seth Hallem, and Dawson~R. Engler.
\newblock An empirical study of operating system errors.
\newblock In {\em Proceedings of the 18th {ACM} Symposium on Operating System
  Principles, {SOSP} 2001, Chateau Lake Louise, Banff, Alberta, Canada, October
  21-24, 2001}, pages 73--88, 2001.

\bibitem{Choudhary2017}
Ankit Choudhary, Shan Lu, and Michael Pradel.
\newblock Efficient detection of thread safety violations via coverage-guided
  generation of concurrent tests.
\newblock In {\em Proceedings of the 39th International Conference on Software
  Engineering, {ICSE} 2017, Buenos Aires, Argentina, May 20-28, 2017}, pages
  266--277, 2017.

\bibitem{Cowan2001}
Crispin Cowan, Steve Beattie, Chris Wright, and Greg Kroah{-}Hartman.
\newblock Raceguard: Kernel protection from temporary file race
  vulnerabilities.
\newblock In {\em 10th {USENIX} Security Symposium, August 13-17, 2001,
  Washington, D.C., {USA}}, 2001.

\bibitem{CVE-2009-1961}
CVE.
\newblock An example of deadlock {CVE-2009-1961} ,
  \url{https://www.cvedetails.com/cve/CVE-2009-1961/}, 2009.

\bibitem{CVE-2016-5195}
CVE.
\newblock The "dirty cow" vulnerability {CVE-2016-5195} ,
  \url{https://www.cvedetails.com/cve/CVE-2016-5195/}, 2016.

\bibitem{CVE-2017-1000380}
CVE.
\newblock {CVE-2017-1000380},
  \url{https://www.cvedetails.com/cve/CVE-2017-1000380/}, 2017.

\bibitem{CVE-2017-10661}
CVE.
\newblock {CVE-2017-10661},
  \url{https://www.cvedetails.com/cve/CVE-2016-5195/}, 2017.

\bibitem{CVE-Details}
CVE.
\newblock {CVE Details}. \url{https://www.cvedetails.com/}, 2018.

\bibitem{CVE-2016-2545}
CVE.
\newblock Race condition in alsa {CVE-2016-2545} ,
  \url{https://www.cvedetails.com/cve/CVE-2016-2545}, 2018.

\bibitem{CVE-2018-1000004}
CVE.
\newblock Race condition in sound system {CVE-2018-1000004} ,
  \url{https://seclists.org/oss-sec/2018/q1/51}, 2018.

\bibitem{Dean2004}
Drew Dean and Alan~J. Hu.
\newblock Fixing races for fun and profit: How to use access(2).
\newblock In {\em Proceedings of the 13th {USENIX} Security Symposium, August
  9-13, 2004, San Diego, CA, {USA}}, pages 195--206, 2004.

\bibitem{Engler2003}
Dawson~R. Engler and Ken Ashcraft.
\newblock Racerx: effective, static detection of race conditions and deadlocks.
\newblock In {\em Proceedings of the 19th {ACM} Symposium on Operating Systems
  Principles 2003, {SOSP} 2003, Bolton Landing, NY, USA, October 19-22, 2003},
  pages 237--252, 2003.

\bibitem{EricksonMBO10}
John Erickson, Madanlal Musuvathi, Sebastian Burckhardt, and Kirk Olynyk.
\newblock Effective data-race detection for the kernel.
\newblock In {\em 9th {USENIX} Symposium on Operating Systems Design and
  Implementation, {OSDI} 2010, October 4-6, 2010, Vancouver, BC, Canada,
  Proceedings}, pages 151--162, 2010.

\bibitem{Fu2018}
Haojie Fu, Zan Wang, Xiang Chen, and Xiangyu Fan.
\newblock A systematic survey on automated concurrency bug detection, exposing,
  avoidance, and fixing techniques.
\newblock {\em Software Quality Journal}, 26(3):855--889, 2018.

\bibitem{Guo2018}
Shengjian Guo, Meng Wu, and Chao Wang.
\newblock Adversarial symbolic execution for detecting concurrency-related
  cache timing leaks.
\newblock In {\em Proceedings of the 2018 {ACM} Joint Meeting on European
  Software Engineering Conference and Symposium on the Foundations of Software
  Engineering, {ESEC/SIGSOFT} {FSE} 2018, Lake Buena Vista, FL, USA, November
  04-09, 2018}, pages 377--388, 2018.

\bibitem{GuoCY17}
Yu~Guo, Yan Cai, and Zijiang Yang.
\newblock {AtexRace}: across thread and execution sampling for in-house race
  detection.
\newblock In {\em Proceedings of the 2017 11th Joint Meeting on Foundations of
  Software Engineering, {ESEC/FSE} 2017, Paderborn, Germany, September 4-8,
  2017}, pages 315--325, 2017.

\bibitem{HuangZhang2016}
Jeff Huang and Charles Zhang.
\newblock Debugging concurrent software: Advances and challenges.
\newblock {\em Journal of Computer Science and Technology}, 31:861--868, 09
  2016.

\bibitem{HuangZD13}
Jeff Huang, Charles Zhang, and Julian Dolby.
\newblock {CLAP:} recording local executions to reproduce concurrency failures.
\newblock In {\em {ACM} {SIGPLAN} Conference on Programming Language Design and
  Implementation, {PLDI} '13, Seattle, WA, USA, June 16-19, 2013}, pages
  141--152, 2013.

\bibitem{Jin2012}
Guoliang Jin, Wei Zhang, and Dongdong Deng.
\newblock Automated concurrency-bug fixing.
\newblock In {\em 10th {USENIX} Symposium on Operating Systems Design and
  Implementation, {OSDI} 2012, Hollywood, CA, USA, October 8-10, 2012}, pages
  221--236, 2012.

\bibitem{Kasikci2012}
Baris Kasikci, Cristian Zamfir, and George Candea.
\newblock Data races vs. data race bugs: telling the difference with portend.
\newblock In {\em Proceedings of the 17th International Conference on
  Architectural Support for Programming Languages and Operating Systems,
  {ASPLOS} 2012, London, UK, March 3-7, 2012}, pages 185--198, 2012.

\bibitem{Kernelthreadsanitizer}
Linux Kernel.
\newblock Kernelthreadsanitizer, a fast data race detector for the linux
  kernel, 2015.

\bibitem{Khoshnood2015}
Sepideh Khoshnood, Markus Kusano, and Chao Wang.
\newblock {ConcBugAssist}: constraint solving for diagnosis and repair of
  concurrency bugs.
\newblock In {\em Proceedings of the 2015 International Symposium on Software
  Testing and Analysis, {ISSTA} 2015, Baltimore, MD, USA, July 12-17, 2015},
  pages 165--176, 2015.

\bibitem{Kocher2018spectre}
Paul Kocher, Daniel Genkin, Daniel Gruss, Werner Haas, Mike Hamburg, Moritz
  Lipp, Stefan Mangard, Thomas Prescher, Michael Schwarz, and Yuval Yarom.
\newblock Spectre attacks: Exploiting speculative execution.
\newblock {\em ArXiv e-prints}, January 2018.

\bibitem{Lhee2005}
Kyung{-}suk Lhee and Steve~J. Chapin.
\newblock Detection of file-based race conditions.
\newblock {\em Int. J. Inf. Sec.}, 4(1-2):105--119, 2005.

\bibitem{Lin2018}
Huarui Lin, Zan Wang, Shuang Liu, Jun Sun, Dongdi Zhang, and Guangning Wei.
\newblock Pfix: fixing concurrency bugs based on memory access patterns.
\newblock In {\em Proceedings of the 33rd {ACM/IEEE} International Conference
  on Automated Software Engineering, {ASE} 2018, Montpellier, France, September
  3-7, 2018}, pages 589--600, 2018.

\bibitem{atomicity-violation-af-packet}
Linux.
\newblock {Atomicity violation code in af\_packet.c},
  \url{https://github.com/torvalds/linux/commit/d199fab63c11998a602205f7ee7ff7c05c97164b},
  2018.

\bibitem{order-violation-seq-queue}
Linux.
\newblock {Order violation code in seq/seq\_queue.c},
  \url{https://git.kernel.org/pub/scm/linux/kernel/git/torvalds/linux.git/commit/?id=3567eb6af614dac436c4b16a8d426f9faed639b3},
  2018.

\bibitem{Lipp2018meltdown}
Moritz Lipp, Michael Schwarz, Daniel Gruss, Thomas Prescher, Werner Haas,
  Stefan Mangard, Paul Kocher, Daniel Genkin, Yuval Yarom, and Mike Hamburg.
\newblock Meltdown.
\newblock {\em ArXiv e-prints}, January 2018.

\bibitem{LiuHuang2018}
Bozhen Liu and Jeff Huang.
\newblock D4: Fast concurrency debugging with parallel differential analysis.
\newblock In {\em Proceedings of the 39th ACM SIGPLAN Conference on Programming
  Language Design and Implementation}, PLDI 2018, pages 359--373, New York, NY,
  USA, 2018. ACM.

\bibitem{Liu2018}
Changming Liu, Deqing Zou, Peng Luo, Bin~B. Zhu, and Hai Jin.
\newblock A heuristic framework to detect concurrency vulnerabilities.
\newblock In {\em Proceedings of the 34th Annual Computer Security Applications
  Conference, {ACSAC} 2018, San Juan, PR, USA, December 03-07, 2018}, pages
  529--541, 2018.

\bibitem{LiuTZ2014}
Peng Liu, Omer Tripp, and Charles Zhang.
\newblock Grail: Context-aware fixing of concurrency bugs.
\newblock In {\em Proceedings of the 22Nd ACM SIGSOFT International Symposium
  on Foundations of Software Engineering}, FSE 2014, pages 318--329, New York,
  NY, USA, 2014. ACM.

\bibitem{Lu2007}
Shan Lu, Soyeon Park, Chongfeng Hu, Xiao Ma, Weihang Jiang, Zhenmin Li,
  Raluca~A. Popa, and Yuanyuan Zhou.
\newblock {MUVI:} automatically inferring multi-variable access correlations
  and detecting related semantic and concurrency bugs.
\newblock In {\em Proceedings of the 21st {ACM} Symposium on Operating Systems
  Principles 2007, {SOSP} 2007, Stevenson, Washington, USA, October 14-17,
  2007}, pages 103--116, 2007.

\bibitem{Lu2008}
Shan Lu, Soyeon Park, Eunsoo Seo, and Yuanyuan Zhou.
\newblock Learning from mistakes: a comprehensive study on real world
  concurrency bug characteristics.
\newblock In {\em Proceedings of the 13th International Conference on
  Architectural Support for Programming Languages and Operating Systems,
  {ASPLOS} 2008, Seattle, WA, USA, March 1-5, 2008}, pages 329--339, 2008.

\bibitem{Lu2006}
Shan Lu, Joseph Tucek, Feng Qin, and Yuanyuan Zhou.
\newblock {AVIO:} detecting atomicity violations via access interleaving
  invariants.
\newblock In {\em Proceedings of the 12th International Conference on
  Architectural Support for Programming Languages and Operating Systems,
  {ASPLOS} 2006, San Jose, CA, USA, October 21-25, 2006}, pages 37--48, 2006.

\bibitem{Lucia2010}
Brandon Lucia, Luis Ceze, and Karin Strauss.
\newblock {ColorSafe}: architectural support for debugging and dynamically
  avoiding multi-variable atomicity violations.
\newblock In {\em 37th International Symposium on Computer Architecture {(ISCA}
  2010), June 19-23, 2010, Saint-Malo, France}, pages 222--233, 2010.

\bibitem{LuciaDSC08}
Brandon Lucia, Joseph Devietti, Karin Strauss, and Luis Ceze.
\newblock Atom-aid: Detecting and surviving atomicity violations.
\newblock In {\em 35th International Symposium on Computer Architecture {(ISCA}
  2008), June 21-25, 2008, Beijing, China}, pages 277--288, 2008.

\bibitem{NVD}
NIST.
\newblock {National Vulnerability Database}. \url{https://nvd.nist.gov/}, 2018.

\bibitem{Paleari2008}
Roberto Paleari, Davide Marrone, Danilo Bruschi, and Mattia Monga.
\newblock On race vulnerabilities in web applications.
\newblock In {\em Detection of Intrusions and Malware, and Vulnerability
  Assessment, 5th International Conference, {DIMVA} 2008, Paris, France, July
  10-11, 2008. Proceedings}, pages 126--142, 2008.

\bibitem{Pratikakis2011}
Polyvios Pratikakis, Jeffrey~S. Foster, and Michael Hicks.
\newblock {LOCKSMITH:} practical static race detection for {C}.
\newblock {\em {ACM} Trans. Program. Lang. Syst.}, 33(1):3:1--3:55, 2011.

\bibitem{ruf2000effective}
Erik Ruf.
\newblock Effective synchronization removal for {Java}.
\newblock In {\em ACM SIGPLAN Notices}, volume~35, pages 208--218. ACM, 2000.

\bibitem{RungtaMV09}
Neha Rungta, Eric~G. Mercer, and Willem Visser.
\newblock Efficient testing of concurrent programs with abstraction-guided
  symbolic execution.
\newblock In {\em Model Checking Software, 16th International {SPIN} Workshop,
  Grenoble, France, June 26-28, 2009. Proceedings}, pages 174--191, 2009.

\bibitem{Savage1997}
Stefan Savage, Michael Burrows, Greg Nelson, Patrick Sobalvarro, and Thomas~E.
  Anderson.
\newblock Eraser: {A} dynamic data race detector for multithreaded programs.
\newblock {\em {ACM} Trans. Comput. Syst.}, 15(4):391--411, 1997.

\bibitem{Sen2008}
Koushik Sen.
\newblock Race directed random testing of concurrent programs.
\newblock In {\em Proceedings of the {ACM} {SIGPLAN} 2008 Conference on
  Programming Language Design and Implementation, Tucson, AZ, USA, June 7-13,
  2008}, pages 11--21, 2008.

\bibitem{Shacham2007}
Hovav Shacham.
\newblock The geometry of innocent flesh on the bone: Return-into-libc without
  function calls (on the x86).
\newblock In {\em Proceedings of the 14th ACM Conference on Computer and
  Communications Security}, CCS '07, pages 552--561, New York, NY, USA, 2007.
  ACM.

\bibitem{Shi2010}
Yao Shi, Soyeon Park, Zuoning Yin, Shan Lu, Yuanyuan Zhou, Wenguang Chen, and
  Weimin Zheng.
\newblock Do {I} use the wrong definition?: Defuse: definition-use invariants
  for detecting concurrency and sequential bugs.
\newblock In {\em Proceedings of the 25th Annual {ACM} {SIGPLAN} Conference on
  Object-Oriented Programming, Systems, Languages, and Applications, {OOPSLA}
  2010, October 17-21, 2010, Reno/Tahoe, Nevada, {USA}}, pages 160--174, 2010.

\bibitem{Sinha2010}
Nishant Sinha and Chao Wang.
\newblock Staged concurrent program analysis.
\newblock In {\em Proceedings of the 18th {ACM} {SIGSOFT} International
  Symposium on Foundations of Software Engineering, 2010, Santa Fe, NM, USA,
  November 7-11, 2010}, pages 47--56, 2010.

\bibitem{Stender2008}
Scott Stender and Alexander Vidergar.
\newblock Concurrency attacks in web applications.
\newblock Technical report, Red Hat, 2008.

\bibitem{TanZP11}
Lin Tan, Yuanyuan Zhou, and Yoann Padioleau.
\newblock {aComment}: mining annotations from comments and code to detect
  interrupt related concurrency bugs.
\newblock In {\em Proceedings of the 33rd International Conference on Software
  Engineering, {ICSE} 2011, Waikiki, Honolulu , HI, USA, May 21-28, 2011},
  pages 11--20, 2011.

\bibitem{kernel}
L.~Torvalds.
\newblock Linux kernel git repository.
\newblock \url{git.kernel. org/pub/scm/linux/kernel/git/torvalds/linux.git},
  2018.

\bibitem{TuLSZ19}
Tengfei Tu, Xiaoyu Liu, Linhai Song, and Yiying Zhang.
\newblock Understanding real-world concurrency bugs in go.
\newblock In {\em Proceedings of the Twenty-Fourth International Conference on
  Architectural Support for Programming Languages and Operating Systems,
  {ASPLOS} 2019, Providence, RI, USA, April 13-17, 2019}, pages 865--878, 2019.

\bibitem{Wang11}
Chao Wang, Mahmoud Said, and Aarti Gupta.
\newblock Coverage guided systematic concurrency testing.
\newblock In {\em International Conference on Software Engineering}, pages
  221--230, 2011.

\bibitem{Wang2017}
Pengfei Wang, Jens Krinke, Kai Lu, Gen Li, and Steve Dodier{-}Lazaro.
\newblock How double-fetch situations turn into double-fetch vulnerabilities:
  {A} study of double fetches in the linux kernel.
\newblock In {\em 26th {USENIX} Security Symposium, {USENIX} Security 2017,
  Vancouver, BC, Canada, August 16-18, 2017.}, pages 1--16, 2017.

\bibitem{Warszawski2017}
Todd Warszawski and Peter Bailis.
\newblock Acidrain: Concurrency-related attacks on database-backed web
  applications.
\newblock In {\em Proceedings of the 2017 {ACM} International Conference on
  Management of Data, {SIGMOD} Conference 2017, Chicago, IL, USA, May 14-19,
  2017}, pages 5--20, 2017.

\bibitem{Watson2007}
Robert N.~M. Watson.
\newblock Exploiting concurrency vulnerabilities in system call wrappers.
\newblock In {\em First {USENIX} Workshop on Offensive Technologies, {WOOT}
  '07, Boston, MA, USA, August 6, 2007}, 2007.

\bibitem{Wei2005}
Jinpeng Wei and Calton Pu.
\newblock {TOCTTOU} vulnerabilities in unix-style file systems: An anatomical
  study.
\newblock In {\em Proceedings of the {FAST} '05 Conference on File and Storage
  Technologies, December 13-16, 2005, San Francisco, California, {USA}}, 2005.

\bibitem{WuW19}
Meng Wu and Chao Wang.
\newblock Abstract interpretation under speculative execution.
\newblock In {\em Proceedings of the 40th ACM SIGPLAN Conference on Programming
  Language Design and Implementation}, 2019.

\bibitem{Wu2015}
Zhendong Wu, Kai Lu, Xiaoping Wang, and Xu~Zhou.
\newblock Collaborative technique for concurrency bug detection.
\newblock {\em International Journal of Parallel Programming}, 43(2):260--285,
  2015.

\bibitem{Yang2012}
Junfeng Yang, Ang Cui, Salvatore~J. Stolfo, and Simha Sethumadhavan.
\newblock Concurrency attacks.
\newblock In {\em 4th {USENIX} Workshop on Hot Topics in Parallelism,
  HotPar'12, Berkeley, CA, USA, June 7-8, 2012}, 2012.

\bibitem{Yu2018-1}
Tingting Yu, Zunchen Huang, and Chao Wang.
\newblock Contesa: Directed test suite augmentation for concurrent software.
\newblock {\em IEEE Transactions on Software Engineering}, 07 2018.

\bibitem{Yu2018-2}
Tingting Yu, Wei Wen, Xue Han, and Jane Hayes.
\newblock Conpredictor: Concurrency defect prediction in real-world
  applications.
\newblock {\em IEEE Transactions on Software Engineering}, pages 1--1, 2018.

\bibitem{Yu2005}
Yuan Yu, Tom Rodeheffer, and Wei Chen.
\newblock Racetrack: efficient detection of data race conditions via adaptive
  tracking.
\newblock In {\em Proceedings of the 20th {ACM} Symposium on Operating Systems
  Principles 2005, {SOSP} 2005, Brighton, UK, October 23-26, 2005}, pages
  221--234, 2005.

\bibitem{Zhang2011ConSeq}
Wei Zhang, Junghee Lim, Ramya Olichandran, Joel Scherpelz, Guoliang Jin, Shan
  Lu, and Thomas~W. Reps.
\newblock Conseq: detecting concurrency bugs through sequential errors.
\newblock In {\em Proceedings of the 16th International Conference on
  Architectural Support for Programming Languages and Operating Systems,
  {ASPLOS} 2011, Newport Beach, CA, USA, March 5-11, 2011}, pages 251--264,
  2011.

\bibitem{Zhang2010}
Wei Zhang, Chong Sun, and Shan Lu.
\newblock Conmem: detecting severe concurrency bugs through an effect-oriented
  approach.
\newblock In {\em Proceedings of the 15th International Conference on
  Architectural Support for Programming Languages and Operating Systems,
  {ASPLOS} 2010, Pittsburgh, Pennsylvania, USA, March 13-17, 2010}, pages
  179--192, 2010.

\bibitem{Zhao2018}
Shixiong Zhao, Rui Gu, Haoran Qiu, Tsz~On Li, Yuexuan Wang, Heming Cui, and
  Junfeng Yang.
\newblock {OWL:} understanding and detecting concurrency attacks.
\newblock In {\em 48th Annual {IEEE/IFIP} International Conference on
  Dependable Systems and Networks, {DSN} 2018, Luxembourg City, Luxembourg,
  June 25-28, 2018}, pages 219--230, 2018.

\bibitem{ZhouSLCL17}
Jinpeng Zhou, Sam Silvestro, Hongyu Liu, Yan Cai, and Tongping Liu.
\newblock {UNDEAD:} detecting and preventing deadlocks in production software.
\newblock In {\em Proceedings of the 32nd {IEEE/ACM} International Conference
  on Automated Software Engineering, {ASE} 2017, Urbana, IL, USA, October 30 -
  November 03, 2017}, pages 729--740, 2017.

\end{thebibliography}

\end{document}